%% file: paper.tex
\documentclass[twocolumn,showpacs,aps,prd,superscriptaddress]{revtex4}
\usepackage{graphicx}
\usepackage{dcolumn}
\usepackage{amsmath}
\usepackage{epsfig}    
\usepackage{rotating}
\usepackage{color}

\input pubboard/babarsym

\long\def\ignore#1{\relax}

\newcommand{\BABARPubYear}    {06}
\newcommand{\BABARPubNumber}  {015}
          
\newcommand{\SLACPubNumber} {11799}
\newcommand{\LANLNumber} {0604007}

\newcommand{\comment}[1]{}

\def\K {\ensuremath{K}\xspace}
\def\mKpi {\ensuremath{m_{K\pi}}\xspace}

\def\Kmaybestar {\ensuremath{K^{(*)}\xspace}}

\def\emu {\ensuremath{e\mu}\xspace}

\def\mkpi {\ensuremath{m_{\kaon\pi}}\xspace}

\def\modekavgll {\ensuremath{B\to K\ellell}\xspace}

\def\modekstee {\ensuremath{B\to K^*e^+e^-}\xspace}
\def\modekstmm {\ensuremath{B\to K^*\mu^+\mu^-}\xspace}
\def\modekee {\ensuremath{B^+\rightarrow K^+\epem}\xspace}

\def\modekavgem {\ensuremath{B\rightarrow K\emu}\xspace}
\def\modekstem {\ensuremath{B\rightarrow K^{*}\emu}\xspace}

\def\modekll {\ensuremath{B^+\rightarrow K^+\ellell}\xspace}
\def\modekstll {\ensuremath{B\rightarrow K^*\ellell}\xspace}

\def\modeksll {\ensuremath{B^0\rightarrow K^0_{\scriptscriptstyle S}\ellell}\xspace}

\def\modekstkll {\ensuremath{B^0\rightarrow K^{*0}\ellell}\xspace}

\def\modekstksll {\ensuremath{B^+\rightarrow K^{*+}\ellell}\xspace}

\def\kshort {\ensuremath{K^0_S}\xspace}
\def\kstar {\ensuremath{K^*}\xspace}
\def\mes {\ensuremath{m_{ES}}\xspace}
\def\deltaE {\ensuremath{\Delta E}\xspace}
\def\ctl {\ensuremath{\cos \theta^{*}}\xspace}
\def\ctk {\ensuremath{\cos \theta_{K}}\xspace}
\def\ctksq {\ensuremath{\cos^2\theta_{K}}\xspace}
\def\ctlsq {\ensuremath{\cos^2\theta^{*}}\xspace}

\def\statekavgll {\ensuremath{K\ellell}\xspace}
\def\statekavgee {\ensuremath{Ke^+e^-}\xspace}
\def\statekavgmm {\ensuremath{K\mu^+\mu^-}\xspace}
\def\statekstee {\ensuremath{K^*e^+e^-}\xspace}
\def\statekstmm {\ensuremath{K^*\mu^+\mu^-}\xspace}
\def\statekee {\ensuremath{K^+\epem}\xspace}
\def\statekll {\ensuremath{K^+\ellell}\xspace}
\def\statekstll {\ensuremath{K^*\ellell}\xspace}
\def\statekmm {\ensuremath{K^+\mumu}\xspace}
\def\stateksee {\ensuremath{K^0_{\scriptscriptstyle S}\epem}\xspace}

\def\stateksmm {\ensuremath{K^0_{\scriptscriptstyle S}\mumu}\xspace}

\def\statekstkee {\ensuremath{K^{*0}\epem}\xspace}
\def\statekstkll {\ensuremath{K^{*0}\ellell}\xspace}
\def\statekstkmm {\ensuremath{K^{*0}\mumu}\xspace}
\def\statekstksee {\ensuremath{K^{*+}\epem}\xspace}
\def\statekstksll {\ensuremath{K^{*+}\ellell}\xspace}
\def\statekstksmm {\ensuremath{K^{*+}\mumu}\xspace}
\def\statekzee {\ensuremath{K^0\epem}\xspace}
\def\statekzll {\ensuremath{K^0\ellell}\xspace}
\def\statekzmm {\ensuremath{K^0\mumu}\xspace}

\newcommand{\ds}{\displaystyle}

\def\figurebox#1#2#3{%
    \def\arg{#3}%
    \ifx\arg\empty
    {\hfill\vbox{\hsize#2\hrule\hbox to #2{\vrule\hfill\vbox to #1{\hsize#2\vfill}\vrule}\hrule}\hfill}%
    \else
    {\hfill\epsfbox{#3}\hfill}%
    \fi}

\begin{document}

\preprint{\babar-PUB-\BABARPubYear/\BABARPubNumber}
\preprint{SLAC-PUB-\SLACPubNumber}
 
\begin{flushleft}
\babar-PUB-\BABARPubYear/\BABARPubNumber\\
SLAC-PUB-\SLACPubNumber\\
hep-ex/\LANLNumber\\ [10mm]
\end{flushleft}
\title{
{\large \bf \boldmath Measurements of branching fractions, rate asymmetries, and angular distributions in the rare decays $B \rightarrow K\ell^+ \ell^-$ and $B \rightarrow K^*\ell^+ \ell^-$
}
\vskip 10mm
}
\input pubboard/authors_feb2006.tex
\date{\today}

\begin{abstract}
We present measurements of the flavor-changing neutral current
decays $B\to K\ell^+\ell^-$ and  $B\to K^*\ell^+\ell^-$,
where $\ell^+\ell^-$ is either an $e^+e^-$ or $\mu^+\mu^-$ pair.
The data sample comprises $229\times 10^6$ $\FourS\to \BB$ decays 
collected with the \babar\ detector at the \pep2 $e^+e^-$ storage ring.  
Flavor-changing neutral current decays are highly suppressed in the Standard Model 
and their predicted properties could be significantly modified
by new physics at the electroweak scale.
We measure the branching fractions 
${\cal B}(\B \rightarrow \K\ellell) = (0.34\pm 0.07 \pm 0.02)
\times 10^{-6}$,
${\cal B}(\B \rightarrow \Kstar\ellell) = (0.78^{+0.19}_{-0.17}\pm 0.11) \times 10^{-6}$,
the direct $\CP$ asymmetries of these decays, and the relative abundances
of decays to electrons and muons. For two regions in $\ell^+\ell^-$ 
mass, above and below $m_{\jpsi}$, we measure partial branching fractions
and the forward-backward angular asymmetry of the lepton pair.  In these
same regions we also measure the $K^*$ 
longitudinal polarization in $B\to K^*\ell^+\ell^-$ decays.  
Upper limits are obtained for the 
lepton flavor-violating decays \modekavgem and \modekstem.
All measurements are consistent with Standard Model expectations.
\end{abstract}

\maketitle
\par    
   
\section{Introduction} \label{sec:intro}

\begin{figure}[b!]
\begin{center}
\includegraphics[height=3cm]{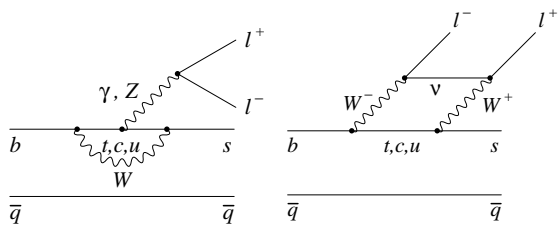}
\caption{Examples of Standard Model Feynman diagrams
for the decays $B\to K^{(*)}\ell^+\ell^-$. For the photon or $Z$ penguin 
diagrams on the left, boson emission can occur on any of the $b$, $t$, $c$, $u$, $s$, 
or $W$ lines.}
\label{fig:PenguinDiagrams}
\end{center}
\end{figure}

The decays $B \rightarrow K^{(*)}\ell^+\ell^-$, where $\ell^+\ell^-$ is either an $e^+e^-$ or $\mu^+\mu^-$ pair and $K^{(*)}$ denotes either a kaon or 
the $K^*(892)$ 
meson, are manifestations of $b \rightarrow s\ell^+\ell^-$ flavor-changing neutral 
currents (FCNC). In the Standard Model (SM), these decays are forbidden 
at tree level and can only occur at greatly suppressed rates through 
higher-order processes. At lowest order, three amplitudes contribute: 
(1) a photon penguin, (2) a $Z$ penguin, and (3) a $W^+W^-$ box diagram 
(Figure ~\ref{fig:PenguinDiagrams}). In all three, a virtual $t$ quark 
contribution dominates, with secondary contributions from virtual 
$c$ and $u$ quarks. Within the Operator Product 
Expansion (OPE) framework, these short-distance contributions  
are typically described in terms of the effective Wilson coefficients 
$C_{7}^{\rm eff}$, $C_{9}^{\rm eff}$, and $C_{10}^{\rm eff}$~\cite{bib:buras}.
Since these decays proceed via weakly-interacting particles with virtual 
energies near the electroweak scale, they provide a promising means to 
search for effects from new interactions entering with amplitudes comparable 
to those of the SM. Such effects are predicted in a wide variety of models
~\cite{bib:chargedhiggs,bib:susy,bib:TheoryA,bib:4g,bib:lq}.

In the SM the $B \rightarrow K\ell^+\ell^-$ branching fraction is predicted to
be roughly $0.4 \times 10^{-6}$, while the $B \rightarrow K^*\ell^+\ell^-$ 
branching fraction is predicted to be about three times larger 
~\cite{bib:TheoryA,bib:TheoryBa,bib:ErratumTheoryBa,bib:TheoryBb,
bib:TheoryBc,bib:TheoryBd,bib:TheoryC}. The $B \rightarrow K^*\ell^+\ell^-$ 
mode receives a significant 
contribution from a pole in the photon penguin amplitude at low values of 
$q^2 \equiv m_{\ell^+\ell^-}^2$, which is not present in \modekavgll decays. 
Due to the lower mass threshold for producing 
an $e^+e^-$ pair, this enhances the $K^*e^+e^-$ final state relative to 
the $K^*\mu^+\mu^-$ state. Currently, theoretical predictions of the
branching fractions have associated uncertainties of about $30 \%$ 
due to form factors that model the hadronic effects in the $B \rightarrow K$ 
or $B \rightarrow K^*$ transition. Previous experimental measurements of the
branching fractions are consistent with the range of theoretical predictions, 
with experimental uncertainties comparable in size to the theoretical 
uncertainties ~\cite{bib:babarprl03,bib:belleprl03}.

With larger datasets, it becomes possible to measure ratios and asymmetries 
in the rates. These can typically be predicted more reliably than the total 
branching fractions. For example, the direct $\CP$ asymmetry 

$$A_{\CP} \equiv \frac{\Gamma(\overline{B} \rightarrow \overline{K}^{(*)}\ell^+\ell^-) - \Gamma(B \rightarrow K^{(*)}\ell^+\ell^-)}{\Gamma(\overline{B} \rightarrow \overline{K}^{(*)}\ell^+\ell^-) + \Gamma(B \rightarrow K^{(*)}\ell^+\ell^-)}$$ 
\noindent is expected to be vanishingly small in the SM, of order 
$10^{-4}$ in the $B \rightarrow K^*\ell^+\ell^-$ mode~\cite{bib:kruger01}. However 
it could be enhanced by new non-SM weak phases~\cite{bib:krugercp}. The ratio 
$R_{K}$, defined as 

$$R_{K} \equiv \frac{\Gamma(B \rightarrow K\mu^+\mu^-)}{\Gamma(B \rightarrow Ke^+e^-)},$$
\noindent also has a precise 
SM prediction of $R_{K} = 1.0000 \pm 0.0001$~\cite{bib:hiller03}. In 
supersymmetric theories with a large ratio ($\tan \beta$) of vacuum 
expectation values of Higgs doublets, $R_{K}$ can be significantly enhanced. 
This occurs via penguin diagrams in which the $\gamma$ or $\Z$ is replaced 
with a neutral Higgs boson that preferentially couples to the heavier 
muons~\cite{bib:yan}. In $B \rightarrow K^*\ell^+\ell^-$ this ratio is 
modified by the photon pole contribution, thus the SM prediction 
is $R_{K^*} \approx 0.75$ ~\cite{bib:TheoryA} with an estimated uncertainty 
of $0.01$~\cite{bib:hiller03} if the pole region is included, 
or $R_{K^*} \approx 1.0$ if it is excluded~\cite{bib:hiller03}.

Additional sensitivity to non-SM physics arises from the fact that 
$B \rightarrow K^{(*)}\ell^+\ell^-$ transitions are three-body decays 
proceeding through three different electroweak penguin amplitudes, whose 
relative contributions vary as a function of $q^2$. 
Measurements of partial branching fractions and angular distributions as a 
function of the invariant momentum transfer $q^2$ are therefore of particular
interest. The SM predicts a distinctive pattern in the forward-backward 
asymmetry 

\begin{eqnarray}     
A_{FB}(s) & \equiv & {\ds \int_{-1}^1 {\rm d} \cos{\theta} \; {\ds{\rm d}^2 
		     \Gamma (B \rightarrow \Kmaybestar \ell^+\ell^-)\over
                     \ds {\rm d} \cos{\theta} \; {\rm d} s} \; {\rm Sign} (\cos{\theta})    
                     \over
                     \ds {\rm d} \Gamma (B \rightarrow \Kmaybestar \ell^+\ell^-)/ {\rm d} s }\nonumber,
\end{eqnarray}

\noindent where $s \equiv q^2/m_{B}^2$, and $\theta$ is the angle of the 
lepton with respect to the flight direction of the $B$ meson, measured in the 
dilepton rest frame ~\cite{bib:buchalla01}.
In the presence of non-SM physics, the sign and magnitude of this asymmetry 
can be altered 
dramatically~\cite{bib:TheoryBb,bib:TheoryA,bib:kruger01}. In particular,
at high $q^2$, the sign of $A_{FB}$ is sensitive to the sign of the
product of the $C_{9}^{\rm eff}$ and $C_{10}^{\rm eff}$ Wilson coefficients. 
The value of $A_{FB}$ in \modekavgll provides an important check on this 
measurement, as it is expected to result in zero asymmetry for all $q^2$ in 
the SM and many non-SM scenarios. This condition can be violated in models 
in which new operators such as a neutral Higgs penguin contribute 
significantly~\cite{bib:yan}. However even in this case the resulting 
asymmetry is expected to be of order $0.01$ or less in the \modekavgll mode 
for electron or muon final states~\cite{bib:demir02}. 
In addition to $A_{FB}$, in \modekstll the fraction of longitudinal 
polarization $F_{L}$ of the $K^*$ can be measured from the angular 
distribution of its decay products. The value of $F_{L}$ measured at 
low $q^2$ is sensitive to effects from
new left-handed currents with complex phases different from the  
SM, resulting in $C_{7}^{\rm eff} = - C_{7}(\textrm{SM})$,
or effects from new right-handed currents in the photon
penguin amplitude~\cite{bib:krugerf0}. The predicted distributions of 
$A_{FB}(q^2)$ and $F_{L}(q^2)$ are shown for the SM 
and for several non-SM scenarios in Figure~\ref{fig:NewPhysAfbF_{L}}.
The non-SM scenarios correspond to those studied 
in Refs.~\cite{bib:TheoryA,bib:TheoryBb,bib:krugerf0}.

Finally, the lepton flavor-violating decays 
$B \rightarrow K^{(*)}e^{\pm}\mu^{\mp}$ can only occur at rates far below
current experimental sensitivities in the context of the SM with neutrino 
mixing. Observation of these decays would therefore be an indication of 
contributions beyond the SM. For example, such decays are allowed in 
leptoquark models~\cite{bib:lq}.

\begin{figure}[b!]
\begin{center}
\includegraphics[width=1.0\linewidth]{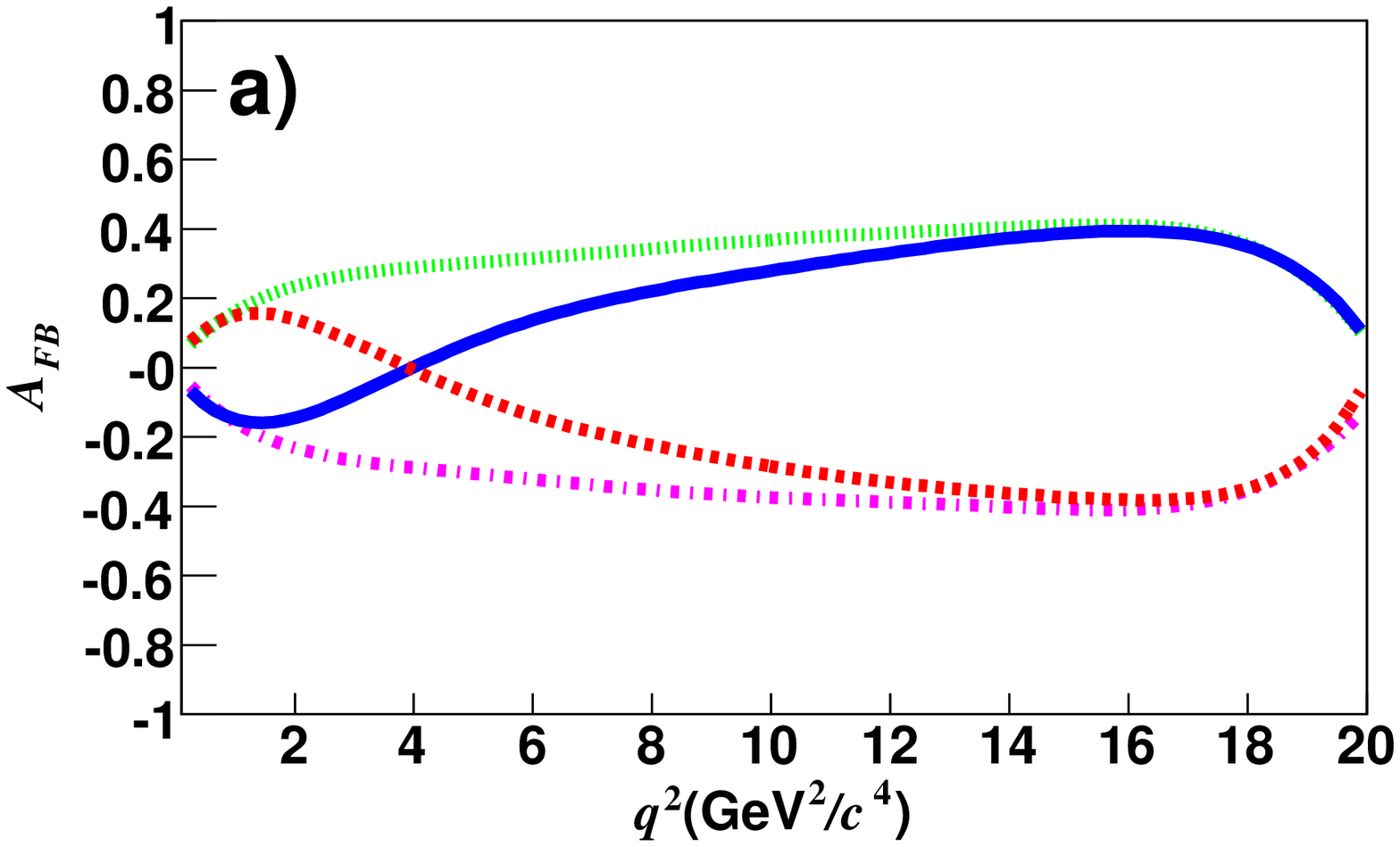}
\includegraphics[width=1.0\linewidth]{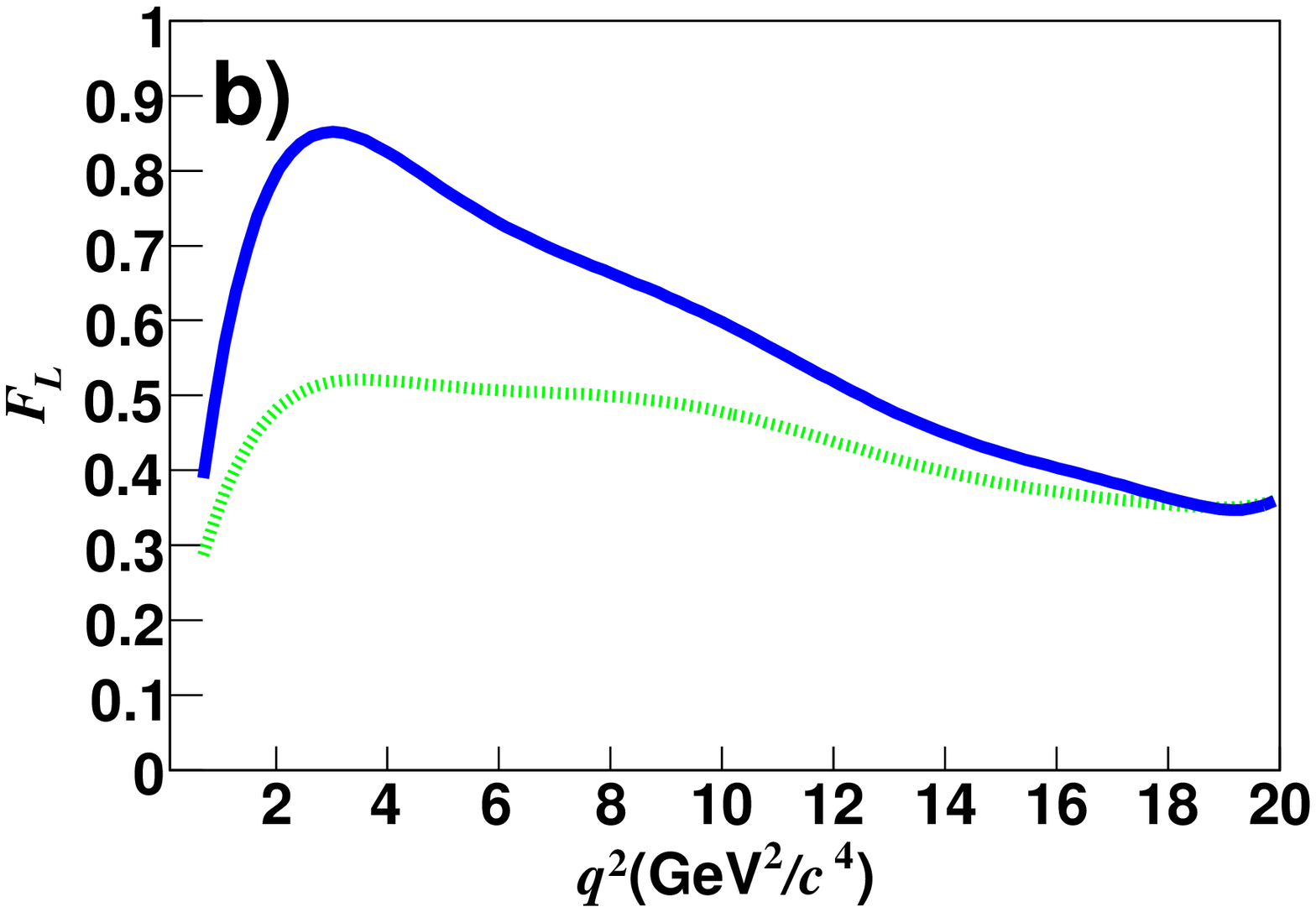}
\caption{Simulated distribution of (a) $A_{FB}$ and (b) $F_{L}$ for 
the decay $B\to K^{*}\ell^+\ell^-$. The points represent the distributions 
assuming the SM (solid lines), $C_{7}^{\rm eff} = -C_{7}(SM)$ 
(dotted lines), $C_{9}^{\rm eff}C_{10}^{\rm eff} = -C_{9}C_{10}(SM)$ 
(dashed lines), and 
$C_{7}^{\rm eff},C_{9}^{\rm eff}C_{10}^{\rm eff} = -C_{7}(SM),-C_{9}C_{10}(SM)$
(dot-dashed lines) generated using the form factor model of 
~\cite{bib:TheoryE}. In the case of $F_{L}$, the two solutions with 
$C_{9}^{\rm eff}C_{10}^{\rm eff} = -C_{9}C_{10}(SM)$ are not displayed; 
they are nearly identical to the two shown.}
\label{fig:NewPhysAfbF_{L}}
\end{center}
\end{figure}

\section{Detector and dataset} \label{sec:detector}

The results presented here are based on data collected with the \babar\ 
detector at the \pep2 asymmetric $e^+e^-$ collider located at the Stanford
Linear Accelerator Center. The dataset comprises 229 million \BB pairs,
corresponding to an integrated luminosity of $208$ \invfb collected on 
the $\Upsilon(4S)$ resonance at a center-of-mass energy of 
$\sqrt{s} = 10.58$ \gev. An additional $12.1$ \invfb of data collected 
at energies $40$ \mev below the nominal on-peak energy is used to study
continuum backgrounds arising from pair production of $u$, $d$, $s$, and 
$c$ quarks.

The \babar\ detector is described in detail in Ref.~\cite{bib:babarNIM}. 
The measurements described in this paper rely primarily on the charged 
particle tracking and identification properties of the detector.
Tracking is provided by a five-layer silicon vertex tracker 
(SVT) and a 40-layer drift chamber (DCH) in a $1.5$-T magnetic field
produced by a superconducting magnet. Low momentum charged hadrons are 
identified by the ionization loss ($dE/dx$) measured in the SVT and DCH, 
and higher momentum hadrons by a ring-imaging detector of internally reflected 
Cherenkov light (DIRC). A CsI(Tl) electromagnetic calorimeter (EMC) provides 
identification of electrons, and detection of photons. The steel in the 
instrumented flux return (IFR) of the superconducting coil is 
interleaved with resistive plate chambers, providing identification of 
muons and neutral hadrons.

\section{Event selection} \label{sec:selection}

\begin{figure*}[]
\begin{center}
\includegraphics[width=0.49\linewidth]{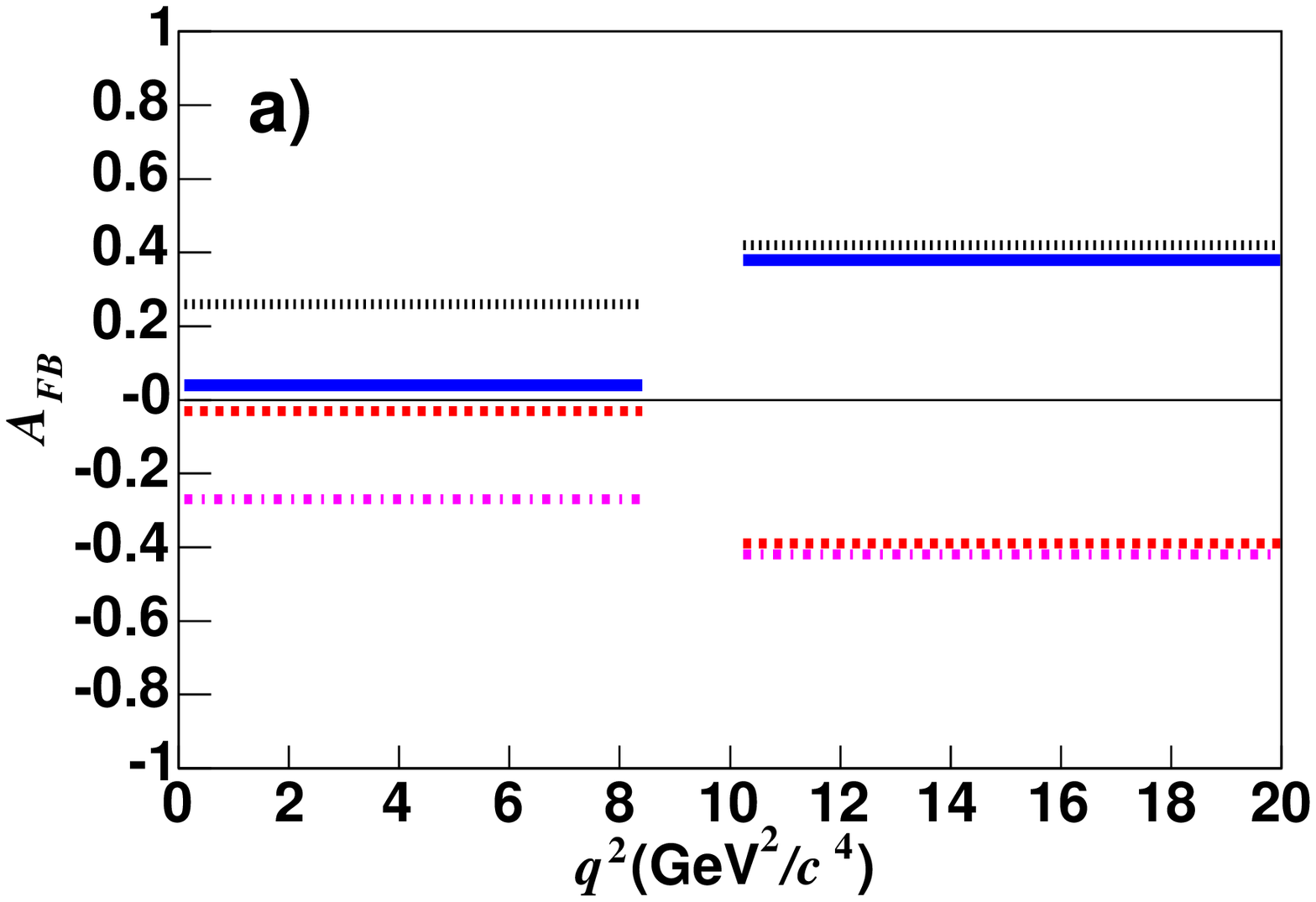}
\includegraphics[width=0.49\linewidth]{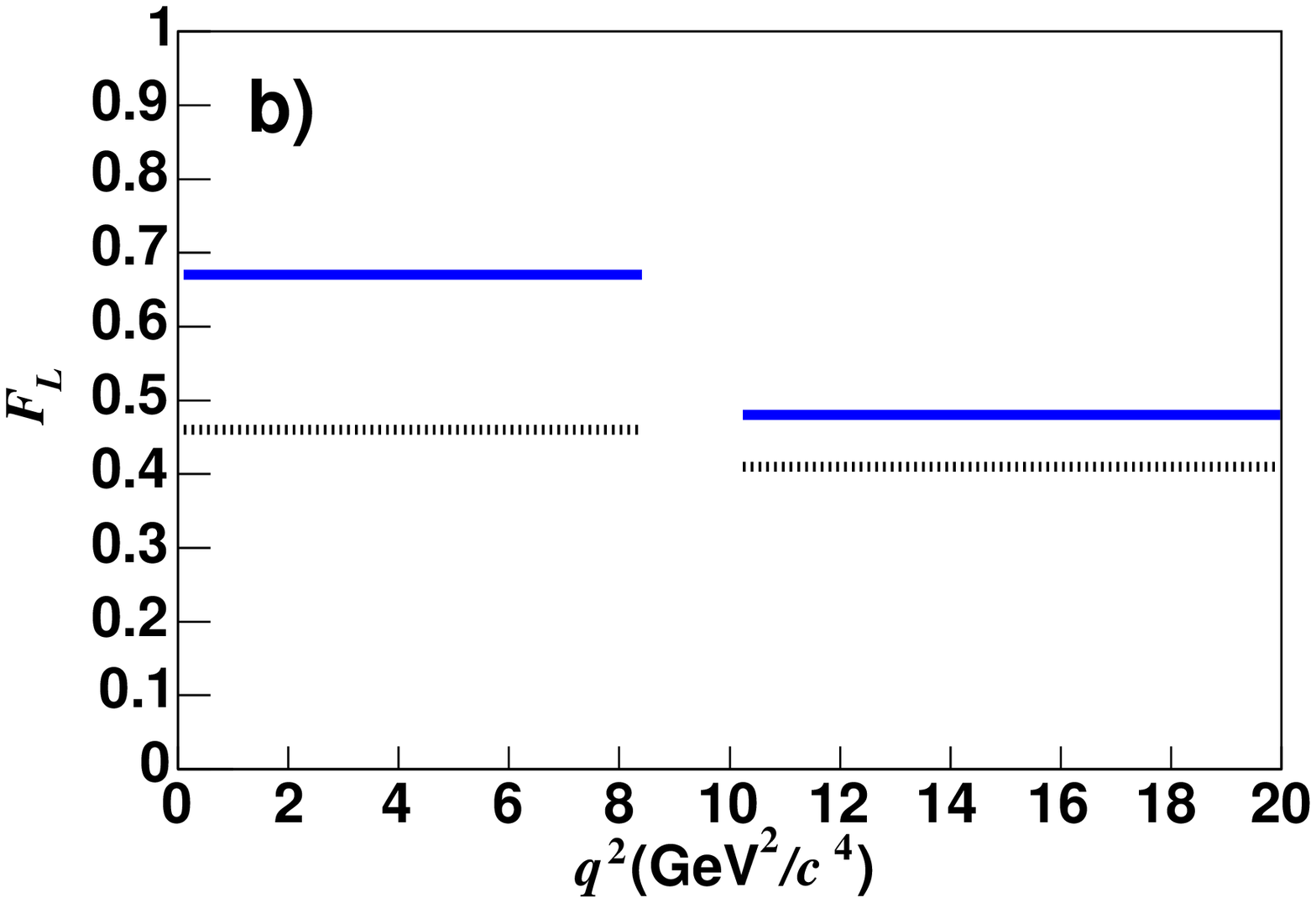}
\caption{Predicted distributions of (a) $A_{FB}(q^{2})$ and 
(b) $F_{L}(q^{2})$ in \modekstll for the two regions of $q^2$ 
considered. The lines represent the predictions of the SM (solid lines), 
$C_{7}^{\rm eff} = -C_{7}(SM)$ (dotted lines), 
$C_{9}^{\rm eff}C_{10}^{\rm eff} = -C_{9}C_{10}(SM)$ (dashed lines), and 
$C_{7}^{\rm eff},C_{9}^{\rm eff}C_{10}^{\rm eff} = -C_{7}(SM),-C_{9}C_{10}(SM)$
(dot-dashed lines) with the form factor model of Ref.~\cite{bib:TheoryE}. In the
case of $F_{L}$, the two solutions with 
$C_{9}^{\rm eff}C_{10}^{\rm eff} = -C_{9}C_{10}(SM)$ are not displayed; 
they are nearly identical to the two shown. }
\label{fig:intasymtheoryplots}
\end{center}
\end{figure*}

We reconstruct signal candidates in eight final states: \modekll, \modeksll, 
\modekstkll, \modekstksll, where 
$K^{*0}\to K^+\pi^-$, $K^{*+}\to K_{\scriptscriptstyle S}^0\pi^+$,
$K_{\scriptscriptstyle S}^0\to\pi^+\pi^-$, and $\ell$ is either an $e$ or $\mu$. Throughout this
paper, charge-conjugate modes are implied.

Electrons are required to have momentum above $0.3$ \gevc and are identified 
using a likelihood ratio combining information from the 
EMC, DIRC, and DCH. Photons that lie in a small angular region around the 
electron direction and have $E > 30$ \mev are combined with electron 
candidates in order to recover bremsstrahlung energy. We suppress backgrounds 
due to photon conversions in the $B \rightarrow Ke^+e^-$ channels by removing
$e^+e^-$ pairs with invariant mass less than 0.03 \gevcc. As there is a 
significant contribution to the $B \rightarrow K^*e^+e^-$ channels from the
pole at low dielectron mass, we preserve acceptance by vetoing conversions in 
these channels only if the conversion radius is outside the inner radius of
the beam pipe.
Muons with momentum $p > 0.7$ \gevc are identified with a neural network 
algorithm using information from the IFR and the EMC.

The performance of the lepton identification algorithms is evaluated using 
high-statistics data control samples. The electron efficiency is determined 
from samples of $e^+e^- \rightarrow e^+e^- \gamma$ events to be approximately 
$91 \%$ over the momentum range considered in this analysis; the pion 
misidentification probability is $< 0.15 \%$, evaluated using control samples 
of pions from $\tau$ and $K_{\scriptscriptstyle S}^0$ decays. The muon efficiency is approximately 
$70 \%$, determined from a sample of $e^+e^- \rightarrow \mu^+\mu^- \gamma$ 
decays; the pion misidentification probability is of order $2 - 3 \%$, 
as determined from $\tau$ decays. These samples are used to correct for any 
discrepancies between data and simulation as a function of momentum, 
polar angle, azimuthal angle, charge, and run period.

Charged kaons are selected by requiring the Cherenkov angle measured
in the DIRC and the track $dE/dx$ to be consistent with the kaon
hypothesis; charged pions are selected by requiring these measurements
to be inconsistent with the kaon hypothesis. \kshort candidates are
constructed from two oppositely charged tracks having an invariant
mass in the range $488.7 < m_{\pi\pi} < 507.3$ \mevcc, a common vertex
displaced from the primary vertex by at least $1$\mm, and a vertex fit
$\chi^2$ probability greater than 0.001. The \kshort mass range corresponds 
to a window of approximately $3 \sigma$ about the nominal \kshort mass. 
Modes that contain a \kstar are required to have a charged $K$ or \kshort 
which, when combined with a charged pion, yields an invariant mass in the 
range $0.7 < m_{K\pi} < 1.1$ \gevcc.

The performance of the charged hadron selection is evaluated using  
control samples of kaons and pions from the decay 
$D^0 \rightarrow K^{-} \pi^{+}$, where the $D^0$ is selected from the 
decay of a $D^*$. The kaon efficiency is determined to be $80 - 97 \%$ over 
the kinematic range relevant to this analysis. The pion misidentification 
probability is $< 3\%$ for momenta less than $3 \gevc$, and increases 
to $\sim 10 \%$ at $5 \gevc$. As with the leptons, these samples are used 
to correct for any discrepancies between the hadron ID performance in data 
and simulation.

Correctly reconstructed $B$ decays will peak in two kinematic variables, \mes 
and \deltaE. For a candidate system of $B$ daughter particles with total 
momentum ${{\bf p}_B}$ in the laboratory frame and energy $E^*_B$ in the 
$\FourS$ center-of-mass (CM) frame, we define 
$\mes = \sqrt{(s/2+ {{\bf p}_0\cdot {\bf p}_B})^2/E^2_0 - {\bf p}^2_B}$
and $\deltaE=E^*_B-\sqrt{s}/2$, where $E_0$ and ${{\bf p}_0}$ are the 
energy and momentum of the $\FourS$ in the laboratory frame, 
and $\sqrt{s}$ is the total CM energy of the $e^+ e^-$ beams.
For signal events, the $m_{\rm ES}$ distribution peaks at the $B$ meson 
mass with resolution $\sigma\approx 2.5\ {\rm MeV}/c^2$.
The $\Delta E$ distribution peaks near zero, with a typical 
width $\sigma \approx$ 18 MeV in the muon channels, and 
$\sigma \approx$ 22 MeV in the electron channels.

$B$ candidates are selected if the reconstructed \mes and \deltaE are in
the ranges $5.00 < \mes < 5.29$ \gevcc and $-0.50 < \deltaE < 0.50$ \gev.
The signal is extracted by performing a multidimensional, unbinned maximum-likelihood fit in the region
$5.20 < \mes < 5.29$ \gevcc and $-0.25 < \deltaE < 0.25$ \gev, which contains 
$100 \%$ of the signal candidates that pass all other selection requirements. 
This region remains blind to our inspection until all selection criteria are 
established. The events in the sideband with $5.00 < \mes < 5.20$ \gevcc, or
$-0.50 < \deltaE < -0.25$ \gev, or $0.25 < \deltaE < 0.50$ \gev are used to 
study the properties of the combinatorial background.

For the measurements of the partial branching fractions, $A_{FB}$, and 
$K^*$ polarization, we subdivide the sample into two
regions of dilepton invariant mass. The first is the region above the pole
and below the $J/\psi$ resonance, $0.1 < q^2 < 8.41 \gev^2/c^4$; 
the second is the region $q^2 > 10.24 \gev^2/c^4$, above the $J/\psi$ 
resonance. The $\psi(2S)$ resonance is explicitly excluded from this upper 
region as described in further detail in Section~\ref{sec:pkgbkg}. The 
lower bound of $0.1 \gev^2/c^4$ in the first region is chosen to remove 
effects from the photon pole in the \modekstee channel. The forward-backward 
asymmetry is extracted in each of these $q^2$ regions from the distribution 
of $\ctl$, which we define as the cosine of the angle between the 
$\ell^-$ ($\ell^+$) and the $B$ ($\overline{B}$) 
meson, measured in the dilepton rest frame. We do not measure $A_{FB}$ in the 
mode \modeksll, in which the flavor of the $B$ meson cannot be directly 
inferred from the $K^0_{\scriptscriptstyle S}$. The $K^*$ polarization is similarly derived from 
the distribution of \ctk, defined as the cosine of the angle between the 
$K$ and the $B$  meson, measured in the $K^*$ rest frame. The predicted 
distributions of $A_{FB}$ and $F_{L}$ integrated over these two $q^2$ ranges 
are shown in 
Figure~\ref{fig:intasymtheoryplots} for both the SM and non-SM scenarios.

\section{Background sources} \label{sec:backgrounds}

\subsection{Combinatorial backgrounds}

Combinatorial backgrounds arise either from the continuum, in which a
($u$, $d$, $s$, or $c$) quark pair is produced, or from \BB events
in which the decay products of the two $B$'s are mis-reconstructed as
a signal candidate. We use the following variables computed in the CM
frame to reject continuum backgrounds: (1) the ratio of second to
zeroth Fox-Wolfram moments~\cite{bib:FoxWolfram}, (2) the angle
between the thrust axis of the $B$ and the remaining particles in the
event, $\theta_{thrust}$, (3) the production angle $\theta_{B}$ of the
$B$ candidate with respect to the beam axis, and (4) the invariant
mass of the kaon-lepton pair with the charge combination expected from a
semileptonic $D$ decay.  The first three variables take advantage of
the characteristic jet-like event shape of continuum backgrounds,
versus the more spherical event shape of \BB events. The fourth
variable is useful for rejecting $c\overline{c}$ events. These
frequently occur through decays such as $D \rightarrow K \ell \nu$,
resulting in a kaon-lepton invariant mass which peaks below that of the
$D$; for signal events the kaon-lepton mass is broadly distributed up to
approximately the $B$ mass. These four variables are combined into a
linear Fisher discriminant~\cite{bib:Fisher}, which is optimized using
samples of simulated signal events and off-resonance data. A separate
Fisher discriminant is used for each of the decay modes considered in
this analysis.

Combinatorial \BB backgrounds are dominated by events with two
semileptonic $B \rightarrow X \ell \nu$ decays.  We discriminate against
these events by constructing a likelihood ratio composed of (1) the
vertex probability of the dilepton pair, (2) the vertex probability of
the $B$ candidate, (3) the angle $\theta_{B}$ as in the Fisher
discriminant, and (4) the total missing energy in the event
$E_{miss}$. Events with two semileptonic decays will contain at least
two neutrinos; therefore the $E_{miss}$ variable is particularly
effective at rejecting these backgrounds. The probability distribution
functions (PDFs) used in the likelihood are derived by fitting
simulated signal events and simulated \BB events in which the
signal decays are removed. We derive a separate likelihood
parameterization for each decay mode.

We select those events that pass an optimal Fisher and \BB likelihood 
requirement, based on the figure of merit $S/\sqrt{S+B}$ for 
the expected number of signal events $S$ and background events $B$. The 
selection is optimized simultaneously for the Fisher and likelihood, and 
is derived separately for each decay mode.

\subsection{Peaking backgrounds}
\label{sec:pkgbkg}

\begin{figure*}[ntb]
\begin{center}
\begin{minipage}{0.4\textwidth}
\includegraphics[height = 6.2 cm]{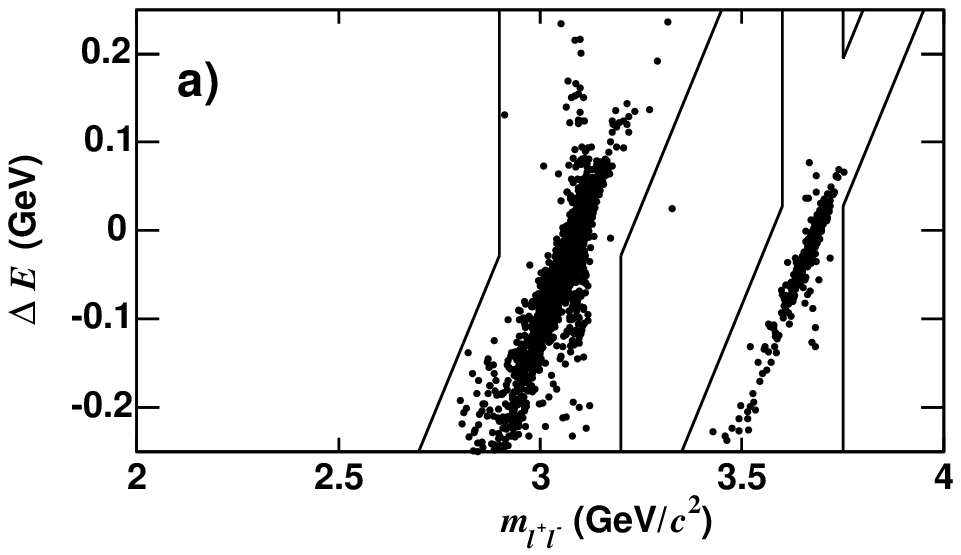}
\end{minipage}
\hfill
\begin{minipage}{0.4\textwidth}
\includegraphics[height = 3.8 cm]{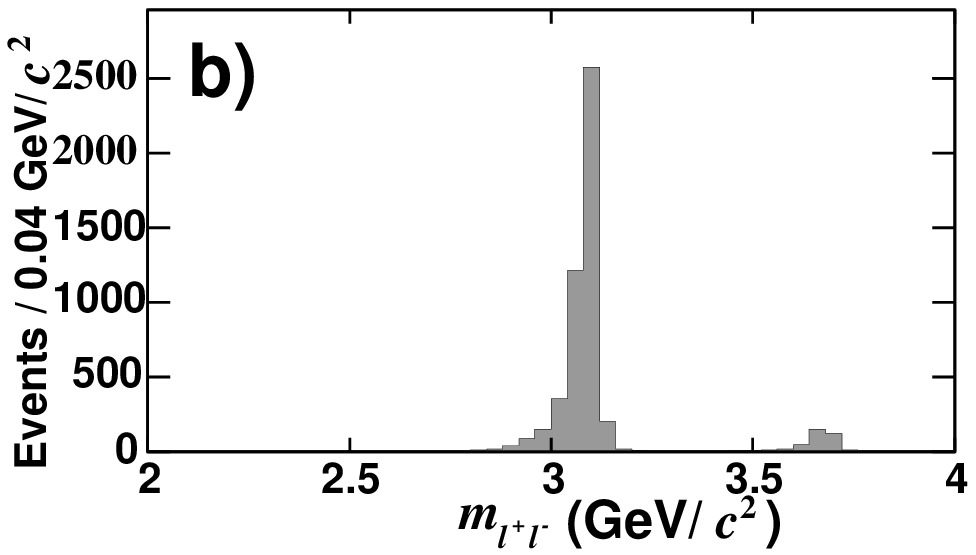}

\includegraphics[height = 3.8 cm]{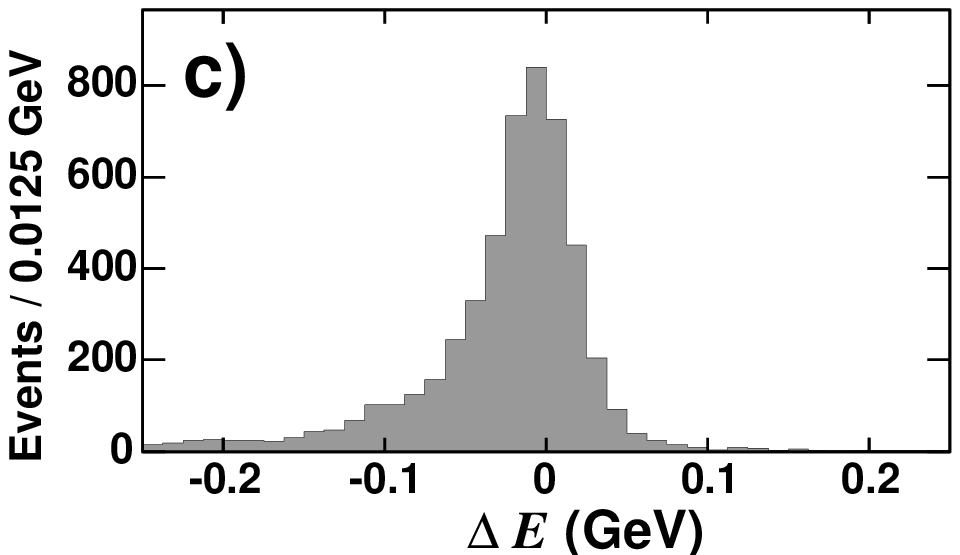}
\end{minipage}

\caption{Charmonium veto regions (a) in the \modekee channel. The points 
are simulated $J/\psi$ and $\psitwos$ events, with abundance equal to the mean 
number expected in 208 \invfb. The projections onto (b) $m_{\ell^+\ell^-}$ 
and (c) $\Delta E$ are shown at right, indicating the high density 
of points at $(m_{\ell^+\ell^-},\Delta E) = (m_{\psi}, 0.0)$.  The vertical 
band corresponds to events where the \jpsi(\psitwos) and \Kp come from 
different \B decays. For $\Delta E < 0$ it also includes events with 
mis-reconstructed $B \rightarrow J/\psi K^*$, $B \rightarrow \psitwos K^*$,  
and non-resonant charmonium decays. The slanted band corresponds to events 
with mis-measured lepton track momentum.}
\label{fig:CharmonVeto}
\end{center}
\end{figure*}

Backgrounds that peak in the \mes and \deltaE variables in the same
manner as the signal are either vetoed, or their rate is estimated
from simulated data or control samples. The largest sources of peaking
backgrounds are $B$ decays to charmonium: $B \rightarrow J/\psi
K^{(*)}$ and $B \rightarrow \psi(2S)K^{(*)}$, where the $J/\psi$ or
$\psi(2S)$ decays to a $\ell^+\ell^-$ pair. We therefore remove events in
which the dilepton invariant mass is consistent with a $J/\psi$ or
$\psi(2S)$, either with or without bremsstrahlung recovery in the
electron channels. In cases where the lepton momentum is mis-measured,
or the bremsstrahlung recovery algorithm fails to find a radiated
photon, the dilepton mass will be shifted from the charmonium mass. In
addition, the measured \deltaE will be shifted away from zero in a
correlated manner. We account for this by constructing a
two-dimensional veto region in the $m_{\ell^+\ell^-}$ vs.  \deltaE plane as
shown in Figure~\ref{fig:CharmonVeto}; the simulated points plotted demonstrate the expected background rejection.
Within the veto region in data we find
approximately 13700 $J/\psi$ events and 1000 $\psi(2S)$ events summed
over all decay modes. These provide a high-statistics control sample
useful for evaluating systematic uncertainties and selection
efficiencies. The residual charmonium background after applying the
veto is estimated from simulation to be between 0.0 and 1.6 events per
decay mode.

Due to the 2-3\% probability for misidentifying pions as muons,
the $B \rightarrow K^{(*)}\mu^+\mu^-$ channels also receive a significant
peaking background contribution from hadronic $B$ decays. The largest of 
these are $B^- \rightarrow D^0\pi^-$ where $D^0 \rightarrow K^-\pi^+$ or 
$D^0 \rightarrow K^{*-}\pi^+$, and $\overline{B^0} \rightarrow D^+\pi^-$ 
where $D^+ \rightarrow \overline{K^{*0}}\pi^+$. These are suppressed by
removing events in which the $K^{(*)}\mu$ invariant mass lies in the range
$1.84 < m_{K^{(*)}\mu} < 1.90$ \gevcc. The remaining hadronic backgrounds
come from charmless decays such as $B \rightarrow K^{(*)}\pi^+\pi^-$,
 $B \rightarrow K^{(*)}K^+\pi^-$, and $B \rightarrow K^{(*)}K^+K^-$.
We measure the peaking background from these processes using data 
control samples of $B \rightarrow K^{(*)}h\mu$ events. These samples are
selected with the same requirements as signal events, except hadron 
identification is required for the hadron candidate $h$ in place
of muon identification. This yields samples of predominantly hadronic
$B$ decays. We then weight each event by the muon misidentification rate
for the hadron divided by its hadron identification efficiency. The 
hadronic peaking background is then extracted by a fit to the \mes 
distribution of these weighted events. This results in a total hadronic 
peaking background measurement of 0.4 - 2.3 events per muon decay channel. 
These backgrounds are suppressed by a factor of approximately 400 in the 
$B \rightarrow K^{(*)}e^+e^-$ channels due to the much lower probability of 
misidentifying pions as electrons.

There is an additional contribution to the peaking backgrounds in the electron
channels from rare two-body decays. These include $B \rightarrow K^*\gamma$ 
with the $\gamma$ converting to an $e^+e^-$ pair in the detector, and 
$B \rightarrow K^{(*)}\pi^0$ or $B \rightarrow K^{(*)}\eta$, where the 
$\pi^0$ or $\eta$ undergoes a Dalitz decay to $e^+e^-\gamma$. These backgrounds
are estimated from simulation to contribute 0.0 - 1.4 events per electron 
decay channel.

\begin{table}[htbp]
\begin{center} 
\caption[Total peaking backgrounds for individual $\Kmaybestar\ellell$ decay modes.]
 {Mean expected peaking backgrounds in 208\invfb, for the individual $\Kmaybestar\ellell$ decay modes after applying all selection requirements.} 
\label{tab:totpkgbkg}
\begin{tabular}{lccc}\hline\hline 
 & All $q^2$ & $0.1 < q^2 < 8.41$ & $q^2 > 10.24$ \\ 
Mode & & ($\gev^2/c^4$) & ($\gev^2/c^4$) \\ 
\hline \vspace{-.1in}\\\vspace{.04in}
\statekee & $0.7 \pm 0.2$ & $0.6 \pm 0.2$ & $0.1 \pm 0.1$\\ \vspace{.04in}
\statekmm & $2.3 \pm 0.5$ & $1.4 \pm 0.4$ & $0.9 \pm 0.1$\\ \vspace{.04in}
\stateksee & $0.01 \pm 0.01$ & $0.01 \pm 0.01$ & $0.0$\\ \vspace{.04in}
\stateksmm & $0.4 \pm 0.1$ & $0.3 \pm 0.1$ & $0.1 \pm 0.04$\\ \vspace{.04in}
\statekstkee & $3.0 \pm 0.6$ & $1.0 \pm 0.5$ & $0.6 \pm 0.2$\\ \vspace{.04in}
\statekstkmm & $1.4 \pm 0.8$ & $0.5 \pm 0.3$ & $0.2 \pm 0.1$\\ \vspace{.04in}
\statekstksee & $0.9 \pm 0.2$ & $0.2 \pm 0.2$ & $0.2 \pm 0.1$\\ \vspace{.04in}
\statekstksmm & $0.6 \pm 0.3$ & $0.2 \pm 0.1$ & $0.2 \pm 0.1$\\\hline\hline
\end{tabular}
\end{center} 

\end{table}

The sum of peaking backgrounds from all sources is summarized in 
Table~\ref{tab:totpkgbkg}. As a function of $q^2$, all of the backgrounds from 
$K^* \gamma$ and $K^{(*)}\pi^0$ are localized in the region 
$0.0 < q^2 < 0.1 \gev^2/c^4$.
Backgrounds from $J/\psi$ and $K^{(*)}\eta$ populate the region 
$0.1 < q^2 < 8.41$ $\gev^2/c^4$, while the $\psi(2S)$ backgrounds contribute 
only to the region $q^2 > 10.24 \gev^2/c^4$. The hadronic backgrounds occupy 
both the $0.1 < q^2 < 8.41$ $\gev^2/c^4$ and $q^2 > 10.24 \gev^2/c^4$ regions.

\section{Yield extraction procedure} \label{sec:fitting}

We extract the signal yield and angular distributions using a 
multidimensional unbinned maximum likelihood fit. For \modekavgll,
the total branching fraction is obtained from a two-dimensional fit 
to \mes and \deltaE. In the \modekstll modes, we add the reconstructed 
$K^*$ mass as a third fit variable. The signal shapes are parameterized 
in both \mes and \deltaE by a Gaussian function plus a radiative tail 
described by an exponential power function. This takes the form

 \[ f(x) \propto
   \left\{ \begin{array}{lrr} 
       \exp(-\frac{(x-\overline{x})^{2}}{2 \sigma^{2}}) &;& (x - \overline{x})/\sigma > \alpha \\ 
	A \times (B - \frac{x - \overline{x}}{\sigma})^{-n} &;&  (x - \overline{x})/\sigma \leq \alpha \\
       \end{array} \right.,\]

\noindent where $A \equiv (\frac{n}{|\alpha|})^{n} \times \exp(-|\alpha|^{2}/2)$ and 
$B \equiv \frac{n}{|\alpha|} - |\alpha|$. The variables $\overline{x}$ and 
$\sigma$ are the Gaussian peak and width, and $\alpha$ and $n$ are 
the point at which the function transitions to the power function 
and the exponent of the power function, respectively. The $m_{\rm ES}$ shape 
parameters $\overline{x}$, $\sigma$, $\alpha$, and $n$ are assumed to 
have a $\Delta E$ dependence of the form $c_0 + c_2(\Delta E)^2$, determined 
empirically from simulation. The mean and width are fixed to the values 
derived by fitting the control sample of vetoed charmonium events. All other 
signal shape parameters are fixed to the values obtained from fits to 
simulated signal events. In the \modekstll mode, the mass of the $K^*$ is 
parameterized with a relativistic Breit-Wigner line shape.

The background is modeled as a sum of terms describing (1) combinatorial 
background; (2) peaking background; (3) cross-feed backgrounds; and, (4) 
in the \modekstll modes, backgrounds that peak in $m_{K\pi}$ at the $K^*$
mass but not in \mes and \deltaE. The combinatorial background is described
by a product of an empirically derived threshold function in \mes, a linear 
term in \deltaE, 
and the product of $\sqrt{m_{K\pi} - m_{K} - m_{\pi}}$ and a quadratic
function of $m_{K\pi}$ for the $K^*$ modes. 
The form of the threshold function used to describe the background in \mes is 
$f(x)\propto x\sqrt{1-x^2}\exp{[-\zeta(1-x^2)]}$, where $\zeta$ is a fit 
parameter and $x=m_{\rm ES}/E_{\rm b}^*$. The peaking background 
component has the same shape as the signal, with normalization fixed
to the estimates of the mean peaking backgrounds (Table~\ref{tab:totpkgbkg}).
The cross-feed component has a floating normalization to describe 
(a) background in $B\to K\ell^+\ell^-$ ($B\to K^{*}\ell^+\ell^-$)
from $B\to K^*\ell^+\ell^-$ ($B\to K^*\pi\ell^+\ell^-$) events with a lost 
pion, and (b) background in $B\to K^*\ell^+\ell^-$ from $B\to K\ell^+\ell^-$ 
events with a randomly added pion. The backgrounds that peak only in
$m_{K\pi}$ are described by the signal shape in $m_{K\pi}$ and the
combinatorial background shape in \mes and \deltaE. The yield of this
term is fixed to $(5 \pm 5)\%$ of the total combinatorial background,
as determined from simulation.
As the shape parameters for term (1) and the normalizations for 
terms (1) and (3) are all free parameters of the fit, 
much of the background uncertainty 
propagates into the statistical uncertainty in the signal yield 
obtained from the fit. 

The $\CP$ asymmetry is also extracted from the fit in the \modekll and
\modekstll channels, where the flavor of the $b$ quark can be inferred 
from the charge of the final state $K^{(*)}$ hadron. As this cannot be done
in the case of \modeksll, we do not measure the $\CP$ asymmetry in that 
mode. The possibility of a non-zero $\CP$ asymmetry in the combinatorial 
background is accounted for by allowing its value to float in the fit. 
The $\CP$ asymmetry of the peaking background is fixed to the value 
expected from the relative composition of background sources.

The partial branching fractions are measured by repeating the fit with 
the sample partitioned into $q^2$ bins. The signal efficiencies 
and peaking backgrounds are recomputed for each region of $q^2$. 
To determine the forward-backward asymmetry and $K^*$ polarization in 
bins of $q^2$, we also utilize fits to the \ctl and \ctk 
angular distributions. We follow the treatment of Ref.~\cite{bib:krugerf0} to 
parameterize the angular distributions for signal. The signal shape 
in \ctk is described by an underlying differential distribution which 
depends on the fraction of longitudinal polarization $F_{L}$ as
$$\frac{1}{\Gamma}\frac{d\Gamma}{d\ctk} = \frac{3}{2} F_L\ctksq + 
\frac{3}{4} (1 - F_L) (1 - \ctksq).$$

\noindent The underlying differential rate for signal in \ctl is 
then described in terms of $F_{L}$ and the forward-backward asymmetry 
term $A_{FB}$ which enters linearly in \ctl: 
\begin{eqnarray}
\frac{1}{\Gamma}\frac{d\Gamma}{d\ctl} = \frac{3}{4} F_L(1 - \ctlsq) +
\frac{3}{8} (1 - F_L) (1 + \ctlsq) + \nonumber\\
A_{FB}\ctl \nonumber.
\end{eqnarray}

\noindent In the \modekll mode, the most general distribution for \ctl with non-zero $A_{FB}$ is given by:
\begin{eqnarray}
\frac{1}{\Gamma}\frac{d\Gamma}{d\ctl} = \frac{3}{4}(1 - F_S)(1 - \ctlsq) + \frac{1}{2}F_S + A_{FB}\ctl \nonumber,
\end{eqnarray}
where $F_S$ is the relative contribution from scalar and pseudoscalar penguin
amplitudes, and $A_{FB}$ arises from the interference of vector and scalar amplitudes~\cite{bib:bobeth01}.  In the Standard Model, both $F_{S}$ and $A_{FB}$
are expected to be negligibly small; their measurement is therefore a null test
sensitive to new physics from scalar or pseudoscalar penguin processes. 

The true angular distributions are altered by detector acceptance 
and efficiency effects. We account for this by multiplying the underlying 
distributions with efficiency functions $\epsilon(\ctl)$
and $\epsilon(\ctk)$ described by a non-parametric histogram PDF obtained
from signal simulations.

The combinatorial background shapes in \ctl and \ctk are described by a 
histogram PDF drawn from control samples in the \mes and \deltaE sidebands.
The angular distribution of the peaking backgrounds are fixed in the fit.
Additional components describing the angular distribution of 
cross-feed events and of mis-reconstructed signal events are included 
as histogram PDFs derived from simulated samples.

In the \modekstll modes we first perform a four-dimensional fit to
\mes, \deltaE, \mKpi, and \ctk to obtain $F_L$.  Due to limited
statistical sensitivity of $F_{L}$ to the \ctl distribution, 
$F_{L}$ is fixed to the value measured from the \ctk distribution
in order to measure $A_{FB}$ from a fit to \mes, \deltaE, \mKpi, and
\ctl.  In the \modekll modes, $A_{FB}$ and $F_{S}$ are simultaneously 
extracted directly from a three-dimensional fit to \mes, \deltaE, and \ctl.

\section{Systematic uncertainties} \label{sec:syst}

\subsection{Branching fractions}

\begin{figure}[b!]
\begin{center}
\includegraphics[scale=0.8]{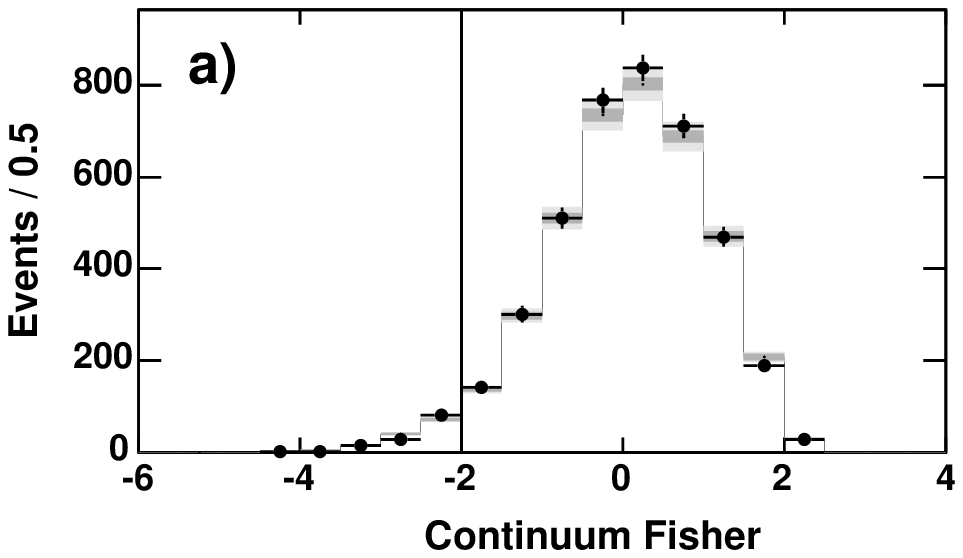}
\includegraphics[scale=0.8]{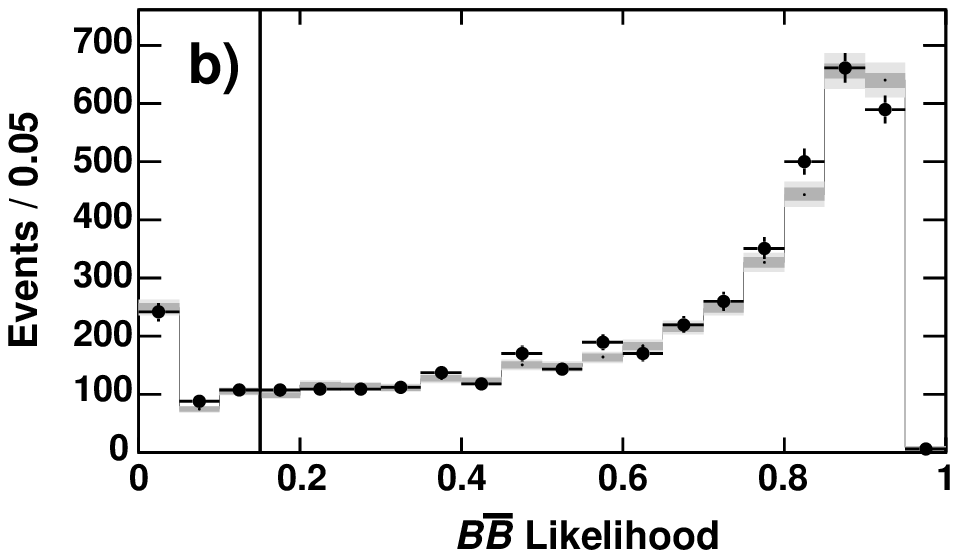}
\caption{Distribution of (a) the Fisher discriminant and (b) the $\BB$ likelihood ratio for $B^+ \rightarrow K^+e^+e^-$ events in the $J/\psi$ veto sample. The points are data; the gray bands are simulated events, with a simulation 
uncertainty given by the band height. The dark gray portion represents the 
uncertainty due to simulation statistics, while the additional 
uncertainty due to the $B \rightarrow J/\psi K^+$ branching fraction 
is represented by the light gray band height. Events to the 
right of the vertical line are selected.}
\label{fig:jpsimultivarplots}
\end{center}
\end{figure}

In evaluating systematic uncertainties in the branching fractions, we
consider both errors that affect the signal efficiency estimate, and
errors arising from the maximum likelihood fit. Sources of
uncertainties that affect the efficiency are: charged-particle
tracking (0.8\% per lepton, 1.4\% per charged hadron),
charged-particle identification (0.5\% per electron pair, 1.3\% per
muon pair, 0.2\% per pion, 0.6\% per kaon), the continuum background
suppression selection (0.3\%--2.2\% depending on the mode), the $\BB$
background suppression selection (0.6\%--2.1\%), $K_{\scriptscriptstyle S}^0$ 
selection
(0.9\%), and signal simulation statistics (0.4\%--0.7\%).  The
estimated number of $\BB$ events in our data sample has an uncertainty
of 1.1\%.  We use the high-statistics sample of events that fail the
charmonium veto to bound the systematic uncertainties associated with
the continuum suppression Fisher discriminant, the $\BB$ likelihood
suppression selection, and charged particle identification. The Fisher
discriminant and $\BB$ likelihood ratio for \modekee are illustrated
in Figure~\ref{fig:jpsimultivarplots} for data and simulation in the
$J/\psi$ control sample. An additional systematic uncertainty in the 
efficiency results from the choice of form factor model, which alters the $q^2$
distribution of the signal. We take this uncertainty to be the maximum
efficiency variation obtained from a set of recent
models~\cite{bib:TheoryD,bib:TheoryE,bib:TheoryBa,bib:ErratumTheoryBa,bib:TheoryBc}; the uncertainty is computed separately for each mode and varies in 
size from 1.1\% to 8.3\%.

Systematic uncertainties on the signal yields obtained from the
maximum-likelihood fit arise from three sources: uncertainties in the
parameters describing the signal shapes, uncertainties in the
combinatorial background shape, and uncertainties in the peaking
backgrounds. The uncertainties in the means and widths of the signal
shapes are obtained by comparing data and simulated data in
$B \rightarrow \jpsi \Kmaybestar$ control samples. For modes with electrons, 
we also vary the fraction of signal events in the tail of the 
$\Delta E$ distribution by varying the
exponent $n$ in the exponential power function. Signal
shape uncertainties are typically 2--4\% of the signal yield.  To
evaluate the uncertainty due to the background shape, we reevaluate
the fit yields with three different parameterizations: (1) an
exponential shape for \DeltaE, (2) a quadratic shape for \DeltaE, and
(3) an \mes background shape parameter $\zeta$ which is linearly correlated
with \DeltaE.  In modes with a $K^*$, we also vary the yield of the 
background component which peaks in \mkpi but not in \mes or \DeltaE 
by 100\% of itself. The induced uncertainty in the signal yield due to the 
background shape is 4--6\% for \modekavgll modes and increases to 
8--12\% for \modekstll modes, where the backgrounds are generally larger.  
Uncertainties in the peaking background induce an uncertainty in the signal 
yields of 2--5\%; this is obtained by varying the expected peaking background 
yield within its $\pm 1 \sigma$ uncertainties. The total systematic 
uncertainty in the fitted signal yield induces a systematic 
uncertainty $\Delta {\cal B}_{\textrm{fit}}$ in the measured branching 
fraction; this uncertainty is shown for each
of the branching fraction fits in Tables~\ref{tab:moderesults} and
~\ref{tab:combinedFits}.

\subsection{$\CP$ asymmetry}

The systematic uncertainties in the measurement of $A_{\CP}$ include
errors due both to detector efficiency effects and to the asymmetry
in the peaking background component. The error associated with the detector 
efficiency is obtained by comparing the value of $A_{\CP}$ measured in the 
charmonium control samples with the expected value of zero; agreement with zero
is obtained with a precision of 1.2\% for \modekll and 2.1\% for
\modekstll.  The  uncertainty due to the peaking background is evaluated by 
varying the expected $\CP$ asymmetry of the peaking backgrounds within their
uncertainties. The possible $\CP$ asymmetry in the charmonium and 
$B \rightarrow K^* \gamma$ peaking backgrounds is highly constrained 
from previous measurements; any asymmetry in the Dalitz decays is suppressed 
by their relatively small contribution to the peaking background. 
In contrast, the hadronic peaking background in the muon modes could 
exhibit a significant $\CP$ asymmetry; this is measured directly from the 
asymmetry of the hadronic control sample described in 
Section~\ref{sec:pkgbkg} with an uncertainty dominated by the statistics 
of the sample.  This induces an uncertainty in the 
measured $A_{\CP}$ of 1\% for \modekll and 2\% for \modekstll.  Other 
systematic uncertainties induced by the fitting procedure, as computed above 
for the branching fraction measurements, are found to be negligible.

\subsection{Angular distributions}

Systematic uncertainties related to the angular distributions of the
efficiency are estimated by comparing the values of $A_{FB}$, $F_{S}$, 
and $F_L$ measured in the relevant charmonium control samples 
with their expected values.
For $B \rightarrow \jpsi K$ and $B \rightarrow \jpsi \Kstar$ we
measure an $A_{FB}$ consistent with zero and with a precision of 0.01
and 0.02, respectively.
  For $B \rightarrow \jpsi \Kstar$, we measure
$F_{L}$ to be consistent with the previous \babar\
measurement~\cite{bib:cos2beta}, with a precision of 0.05.
For $B \rightarrow \jpsi K$ we
measure $F_S$ consistent with zero and with a precision of 0.03. 
 
Further systematic uncertainties are evaluated by repeating the fit
with alternative shapes assumed for the background components: (1) the
shape of mis-reconstructed signal events is fixed instead to the shape
of correctly reconstructed signal, (2) the combinatorial background
shape is drawn from alternative ranges of \mes and \deltaE, and from
the sample of events that fail the \BB likelihood selection, and
(3) the angular distributions of the peaking backgrounds are varied
within their statistical uncertainties.  Systematic uncertainties from
backgrounds induce uncertainties in $F_L$ and $A_{FB}$ of 0.05--0.18,
depending on the relative amount of background, and are the largest 
systematic uncertainty.  $F_S$ is more sensitive to the background shape,
with an induced systematic uncertainty of 0.45.

In the fit to \ctl in the \modekstll decay modes, the value of $F_L$
is fixed to the result obtained from the fit to the \ctk
distribution. This introduces an additional parametric uncertainty of
0.01 on the measured value of $A_{FB}$ , which we evaluate by varying $F_L$ 
within the uncertainty of the measurement.

\section{Results} \label{sec:results}

\subsection{Branching fractions}

\begin{figure}[!tbh]
\begin{center}
\includegraphics[width=1.0\linewidth]{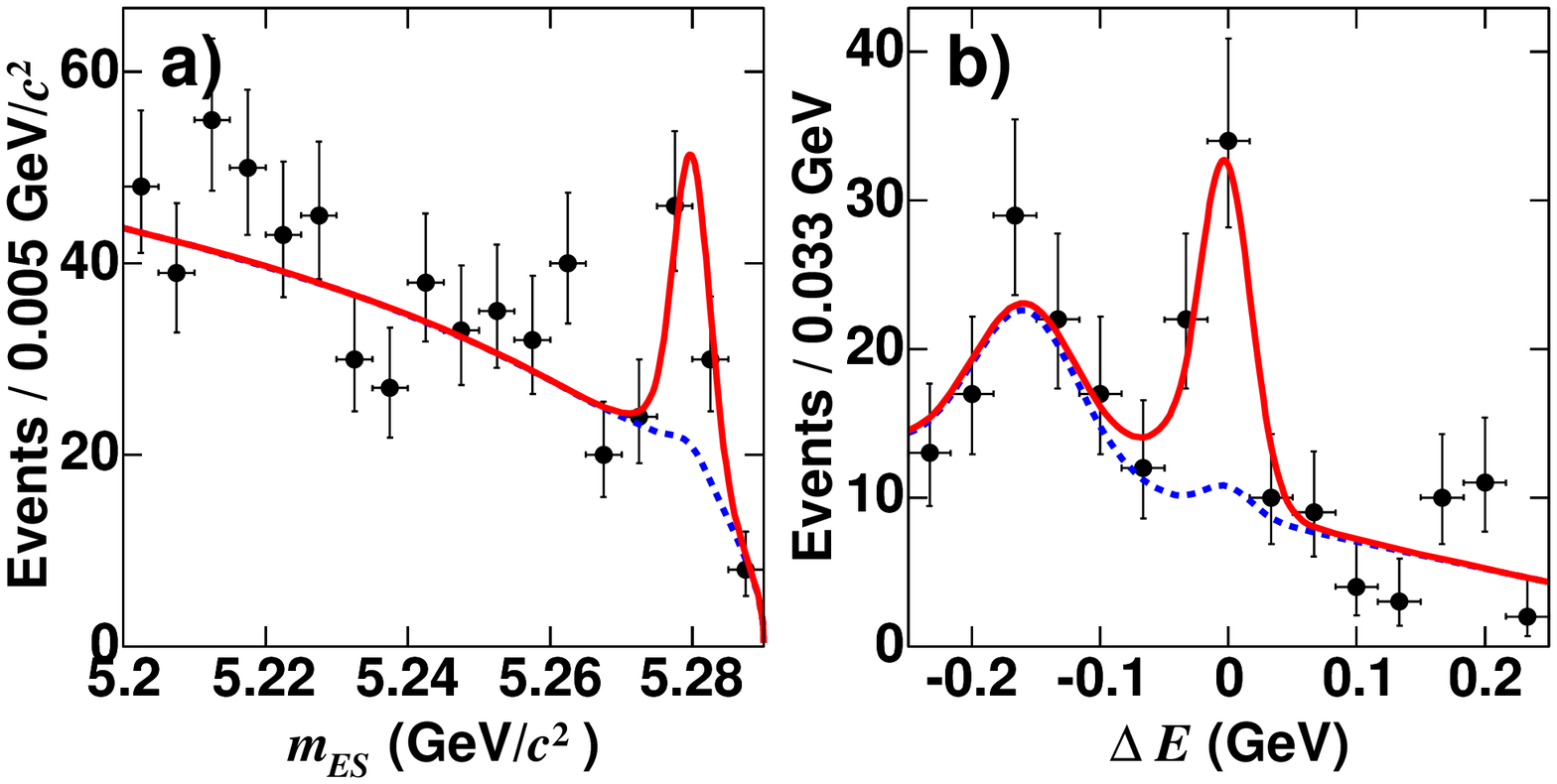}
\caption{
Distributions of the fit variables in $K\ell^+\ell^-$ data (points),
compared with projections of the combined fit (curves): (a) $m_{\rm
ES}$ distribution after requiring $-0.11<\Delta E<0.05\ {\rm GeV}$ and
(b) $\Delta E$ distribution after requiring 
$|m_{\rm ES} - m_{B}| < 6.6\ {\rm MeV}/c^2$.
The solid curve is the sum of all fit components,
including signal; the dashed curve is the sum of all background
components.
}
\label{fig:dataFitKll}
\end{center}
\end{figure}

\begin{figure}[!tbh]
\begin{center}
\includegraphics[width=1.0\linewidth]{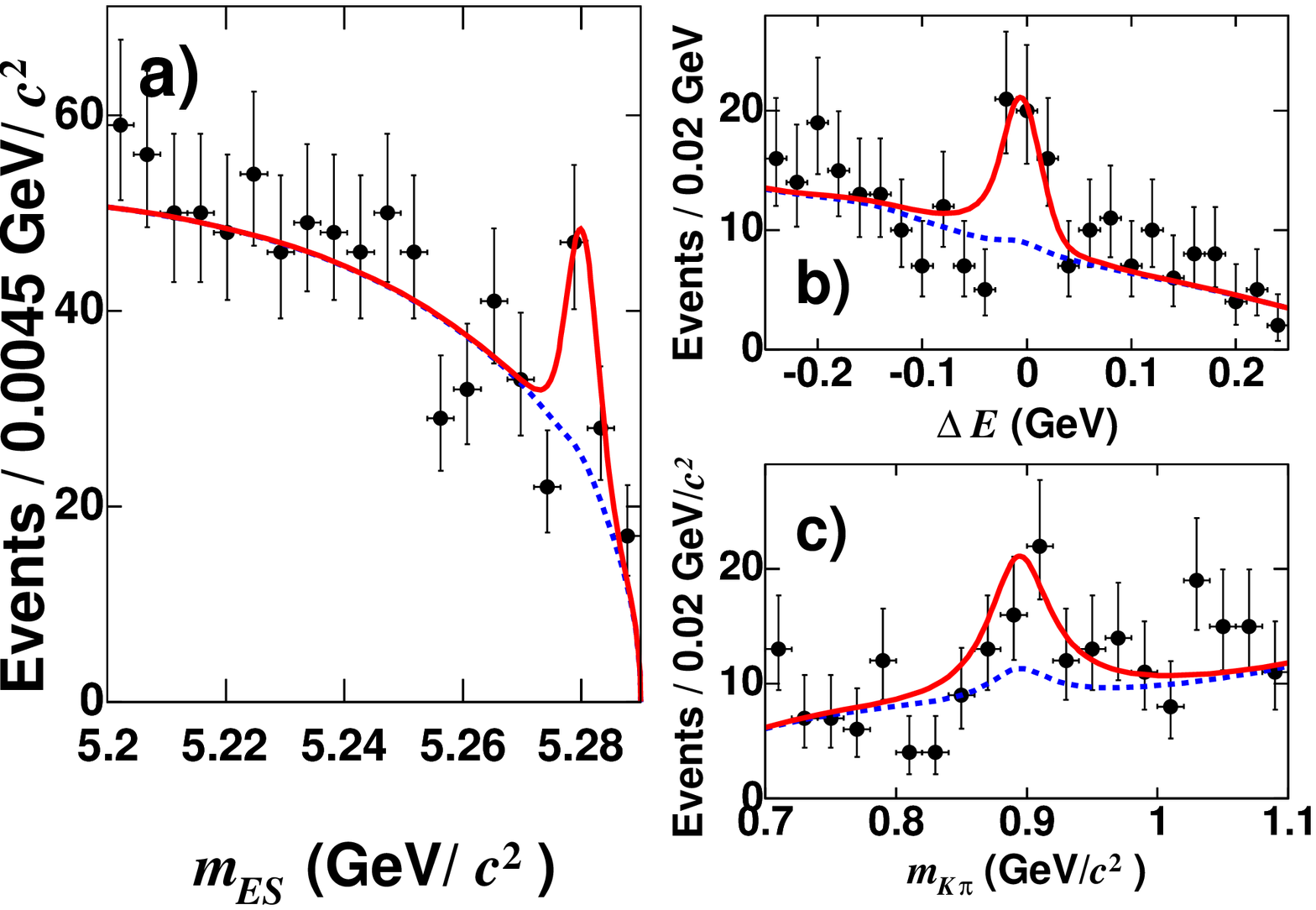}
\caption{
Distributions of the fit variables in $K^*\ell^+\ell^-$ data (points),
compared with projections of the combined fit (curves):
(a) \mes after requiring $-0.11<\Delta E<0.05\ {\rm GeV}$
and $0.817< \mkpi <0.967\ {\rm GeV}/c^2$,
(b) $\Delta E$ after requiring 
$|m_{\rm ES} - m_{B}| < 6.6\ {\rm MeV}/c^2$,
$0.817< \mkpi <0.967\ {\rm GeV}/c^2$, and   
(c) \mkpi after requiring 
$|m_{\rm ES} - m_{B}| < 6.6\ {\rm MeV}/c^2$
and $-0.11<\Delta E<0.05\ {\rm GeV}$. 
The solid curve is the sum of all fit components, including signal; the 
dashed curve is the sum of all background components.
}
\label{fig:dataFitKstll}
\end{center}
\end{figure}

We first perform the fit separately for each of the eight decay modes to
extract the branching fractions integrated over all $q^2$. In the branching 
fraction fits, the efficiency is defined such that the total branching 
fraction includes the estimated signal that is lost due to the 
charmonium vetos.
The results for the individual decay modes are shown in 
Table~\ref{tab:moderesults}. We then perform a combined fit to
the appropriate combinations of modes to extract the 
$B \rightarrow K\ell^+\ell^-$ and $B \rightarrow K^*\ell^+\ell^-$ branching 
fractions. We combine charged and neutral modes by constraining the total 
width ratio $\Gamma(B^0)/\Gamma(B^+)$ to the world average ratio of 
lifetimes $\tau(B^+)/\tau(B^0) = 1.071 \pm 0.009$~\cite{bib:pdglifetime}.
In the $B \rightarrow K^*\ell^+\ell^-$ mode, we add the additional constraint
$\Gamma(B \rightarrow K^*\mu^+\mu^-)/\Gamma(B \rightarrow K^*e^+e^-) = 0.75$
to account for the enhancement due to the pole at low $q^2$ in the electron 
channel~\cite{bib:TheoryA}. The final branching fractions are expressed in 
terms of the $B^0 \rightarrow K^{(*)0}\mu^+\mu^-$ channels. With these 
constraints, we find the lepton-flavor averaged, $B$-charge averaged branching
fractions
$${\cal B}(\modekavgll) = (0.34 \pm 0.07 \pm 0.02) \times 10^{-6},$$
$${\cal B}(\modekstll) = (0.78^{+0.19}_{-0.17} \pm 0.11) \times 10^{-6},$$
where the first error is statistical and the second systematic.
The projections of the data overlayed with the combined fit results are 
shown in Figures~\ref{fig:dataFitKll} and 
~\ref{fig:dataFitKstll}. The signal significance is computed as 
$\sqrt{2 \Delta \ln(\cal{L})}$, where $\Delta \ln(\cal{L})$ is the 
difference between the likelihood of the best fit and that of the null signal 
hypothesis. Systematic uncertainties are incorporated in the significance 
estimate by simultaneously applying all variations that result in a lower 
signal yield before computing the change in likelihood. The significance of 
the signal including statistical and systematic uncertainties is $6.6$ 
standard deviations for the \modekavgll mode and $5.7$ standard deviations 
for the \modekstll mode. The secondary peak in the \DeltaE sideband of 
\modekavgll results from the 
fit component describing events with a lost pion, either from \modekstll 
or from events in which a $b \rightarrow s\ell^+\ell^-$ decay results in a 
$K \pi \ell^+\ell^-$ final state without proceeding through an 
intermediate $K^*$ resonance.  The normalization and mean \DeltaE of this 
component are free parameters in the fit.  
Examination of these events shows that the 
addition of a charged or neutral pion results in a \modekstll or 
$B \rightarrow K \pi \ell^+\ell^-$ signal candidate. Using simulated 
signal decays, we find the effect of these events on the \modekavgll signal 
yield is negligible.

We further perform a set of combined fits with the sample partitioned into 
final states containing muons and electrons, and into charged and neutral 
final states, modifying the constraints as appropriate. The results from 
all such fits are summarized in Table~\ref{tab:combinedFits}.

\begin{table}[h]
\begin{center} 
\caption[Results from fits to the individual $\Kmaybestar\ellell$ decay modes.]
 {Results from fits to the individual $\Kmaybestar\ellell$ decay modes for
all $q^2$. The columns from left are: decay mode, fitted signal yield, signal 
efficiency, relative uncertainty on the branching fraction due to the systematic error 
on the efficiency estimate, systematic error 
from the fit, and the resulting branching fraction (with statistical and 
systematic errors).} 
\label{tab:moderesults}
\begin{tabular}{lD{.}{.}{3.5}ccccc}\hline\hline
& &\multicolumn{1}{c}{$\epsilon$} & \multicolumn{1}{c}{$\Delta\cal B_{\rm eff}$} & \multicolumn{1}{c}{$\Delta \cal B_{\rm fit}$} &\multicolumn{1}{c}{$\cal B$} \\
\multicolumn{1}{c}{Mode} & \multicolumn{1}{c}{Yield} & \multicolumn{1}{c}{$(\%)$} & \multicolumn{1}{c}{$(\%)$} & \multicolumn{1}{c}{$(10^{-6})$} & \multicolumn{1}{c}{$(10^{-6})$}\\\hline \vspace{-.1in}\\\vspace{.04in}
\statekee &25.9^{+7.4}_{-6.5} &$26.6$ &$\pm 3.7$ & $\pm 0.02$ &$0.42^{+0.12}_{-0.11} \pm 0.02$\\ \vspace{.04in}
\statekmm &10.9^{+5.1}_{-4.3} &$15.4$ &$\pm 4.1$ & $\pm 0.03$ &$0.31^{+0.15}_{-0.12} \pm 0.03$\\ \vspace{.04in}
\statekzee &2.4^{+2.8}_{-2.0} &$22.8$ &$\pm 9.6$ & $\pm 0.01$ &$0.13^{+0.16}_{-0.11} \pm 0.02$\\ \vspace{.04in}
\statekzmm &6.3^{+3.6}_{-2.8} &$13.6$ &$\pm 8.3$ & $\pm 0.04$ &$0.59^{+0.33}_{-0.26} \pm 0.07$\\ \vspace{.04in}
\statekstkee &29.4^{+9.5}_{-8.4} &$18.6$ &$\pm 4.9$ & $\pm 0.09$ &$1.04^{+0.33}_{-0.29} \pm 0.11$\\ \vspace{.04in}
\statekstkmm &15.9^{+7.0}_{-5.9} &$11.9$ &$\pm 5.8$ & $\pm 0.11$ &$0.87^{+0.38}_{-0.33} \pm 0.12$\\ \vspace{.04in}
\statekstksee &6.1^{+6.3}_{-5.3} &$15.7$ &$\pm 6.8$ & $\pm 0.37$ &$0.75^{+0.76}_{-0.65} \pm 0.38$\\ \vspace{.04in}
\statekstksmm &4.7^{+4.6}_{-3.4} &$9.3$ &$\pm 7.1$ & $\pm 0.13$ &$0.97^{+0.94}_{-0.69} \pm 0.14$\\\hline\hline
\end{tabular}
\end{center} 
\end{table}

\begin{table}[h]
\begin{center} 
\caption[Results from fits to the combined $\Kmaybestar\ellell$ decay modes.]
 {
Results from fits to combined $\Kmaybestar\ellell$ decay modes for 
all $q^2$. The columns from left are: decay mode combination, 
fitted signal yield,
relative uncertainty on 
the branching fraction due to the systematic error on the efficiency estimate, 
systematic error on the branching fraction introduced by the systematic error 
on the fitted signal yield, and the resulting branching fraction (with 
statistical and systematic errors).  The constraints for each combined fit 
are described in the text. 
} 
\label{tab:combinedFits}
\begin{tabular}{lD{.}{.}{3.5}cccc}\hline\hline
& \multicolumn{1}{c}{Yield} & \multicolumn{1}{c}{$\Delta\cal B_{\rm eff}$} & \multicolumn{1}{c}{$\Delta \cal B_{\rm fit}$} &\multicolumn{1}{c}{$\cal B$}\\
\multicolumn{1}{c}{Mode} & \multicolumn{1}{c}{(events)} & \multicolumn{1}{c}{$(\%)$} & \multicolumn{1}{c}{$(10^{-6})$} & \multicolumn{1}{c}{$(10^{-6})$}\\\hline \vspace{-.1in}\\\vspace{.04in}
\statekavgee  &28.1^{+7.8}_{-7.0}  & $\pm 4.7$ & $\pm 0.02$ &$0.33^{+0.09}_{-0.08} \pm 0.02$\\ \vspace{.04in}
\statekavgmm  &17.3^{+6.2}_{-5.4}  & $\pm 4.8$ & $\pm 0.03$ &$0.35^{+0.13}_{-0.11} \pm 0.03$\\ \vspace{.04in}
\statekll     &36.7^{+8.8}_{-7.9}  & $\pm 3.7$ & $\pm 0.02$ &$0.38^{+0.09}_{-0.08} \pm 0.02$\\ \vspace{.04in}
\statekzll    &8.2^{+4.5}_{-3.6}   & $\pm 9.0$ & $\pm 0.02$ &$0.29^{+0.16}_{-0.13} \pm 0.03$\\ \vspace{.04in}
\statekavgll  &45.5^{+9.8}_{-8.9}  & $\pm 4.6$ & $\pm 0.02$ &$0.34^{+0.07}_{-0.07} \pm 0.02$\\ \vspace{.04in}
\statekstee   &36.2^{+11.2}_{-10.0}& $\pm 5.2$ & $\pm 0.13$ &$0.97^{+0.30}_{-0.27} \pm 0.14$\\ \vspace{.04in}
\statekstmm   &20.7^{+8.1}_{-7.0}  & $\pm 5.9$ & $\pm 0.11$ &$0.88^{+0.35}_{-0.30} \pm 0.12$\\ \vspace{.04in}
\statekstkll  &45.3^{+11.6}_{-10.5}& $\pm 5.0$ & $\pm 0.08$ &$0.81^{+0.21}_{-0.19} \pm 0.09$\\ \vspace{.04in}
\statekstksll &11.5^{+8.0}_{-6.6}  & $\pm 6.6$ & $\pm 0.20$ &$0.73^{+0.50}_{-0.42} \pm 0.21$\\ \vspace{.04in}
\statekstll   &57.1^{+13.7}_{-12.5}& $\pm 5.3$ & $\pm 0.10$ &$0.78^{+0.19}_{-0.17} \pm 0.11$\\ \hline
Pole excluded  & & & & \\ \vspace{.04in}
\statekstee   &23.6^{+9.4}_{-8.3}& $\pm 5.2$ & $\pm 0.11$ &$0.63^{+0.25}_{-0.22} \pm 0.11$ \\ \vspace{.04in}
\statekstmm   &20.7^{+8.1}_{-7.0}  & $\pm 5.9$ & $\pm 0.11$ &$0.88^{+0.34}_{-0.30} \pm 0.12$ \\ \vspace{.04in}
\statekstkll  &34.8^{+10.4}_{-9.3}& $\pm 5.0$ & $\pm 0.10$ &$0.75^{+0.22}_{-0.20} \pm 0.10$ \\ \vspace{.04in}
\statekstksll &9.5^{+7.0}_{-5.7}  & $\pm 6.6$ & $\pm 0.19$ &$0.73^{+0.53}_{-0.44} \pm 0.19$ \\ \vspace{.04in}
\statekstll   &44.3^{+12.2}_{-11.1}& $\pm 5.3$ & $\pm 0.11$ &$0.73^{+0.20}_{-0.18} \pm 0.11$ \\
\hline\hline
\end{tabular}
\end{center} 
\end{table}

If the pole region is removed by requiring $q^2 > 0.1 \gev^2/c^4$, the 
constrained ratio between \modekstmm and \modekstee in the combined fit 
is modified from 0.75 to 1. Repeating the combined fit with this modification, we obtain
$${\cal B}(\modekstll)_{(q^2 > 0.1 \gev^2/c^4)} = (0.73 ^{+0.20}_{-0.18} \pm 0.11) \times 10^{-6}.$$
The results of the combined fits in the various subsamples 
with the pole region removed are shown in Table~\ref{tab:combinedFits}. 
We observe good agreement in the branching fraction obtained in all of the 
subsamples, both with and without the pole region included. The measured 
total rates are consistent with the range of Standard Model rates predicted 
in Ref.~\cite{bib:TheoryA}. The \modekavgll rate is significantly lower than 
the range given by Ref.~\cite{bib:TheoryC}.

From the separate fits to the muon and electron channels integrated over 
all $q^2$, we obtain the ratios
$$R_K= 1.06 \pm 0.48 \pm 0.08,$$
$$R_{K^*}= 0.91 \pm 0.45 \pm 0.06,$$
consistent with the SM predictions of 1.00 and 0.75, 
respectively. If instead the pole region is excluded from the 
$B \rightarrow K^*\ell^+\ell^-$ channels, we find 
$$R_{K^*, (q^2 > 0.1 \gev^2/c^4)} = 1.40 \pm 0.78 \pm 0.10,$$
where this ratio is expected to be 1 in the SM.

\subsection{$\CP$ asymmetry}

From the fit to the combined modes integrated over all $q^2$, we find the 
direct $\CP$ asymmetries
$$A_{\CP}(B^+ \rightarrow K^+\ell^+\ell^-) = -0.07 \pm 0.22 \pm 0.02,$$
$$A_{\CP}(B \rightarrow K^*\ell^+\ell^-) = +0.03 \pm 0.23 \pm 0.03,$$
where the first error is statistical and the second systematic. 
The measured values in both channels are consistent with the SM expectation 
of a negligible direct $\CP$ asymmetry.

\subsection{Partial branching fractions}

\begin{figure*}[ntb]
\begin{center}
\includegraphics[width=0.49\linewidth]{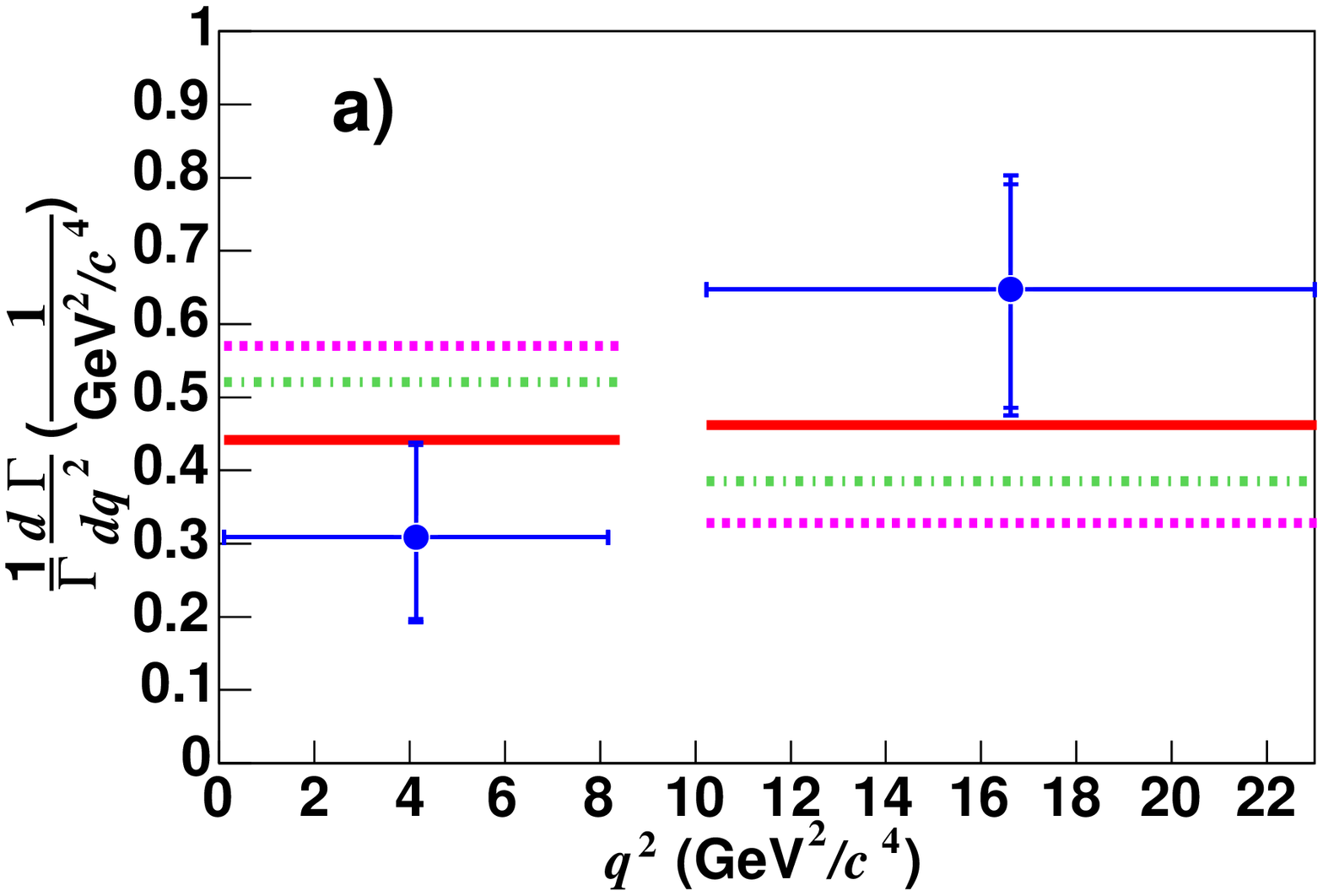}
\includegraphics[width=0.49\linewidth]{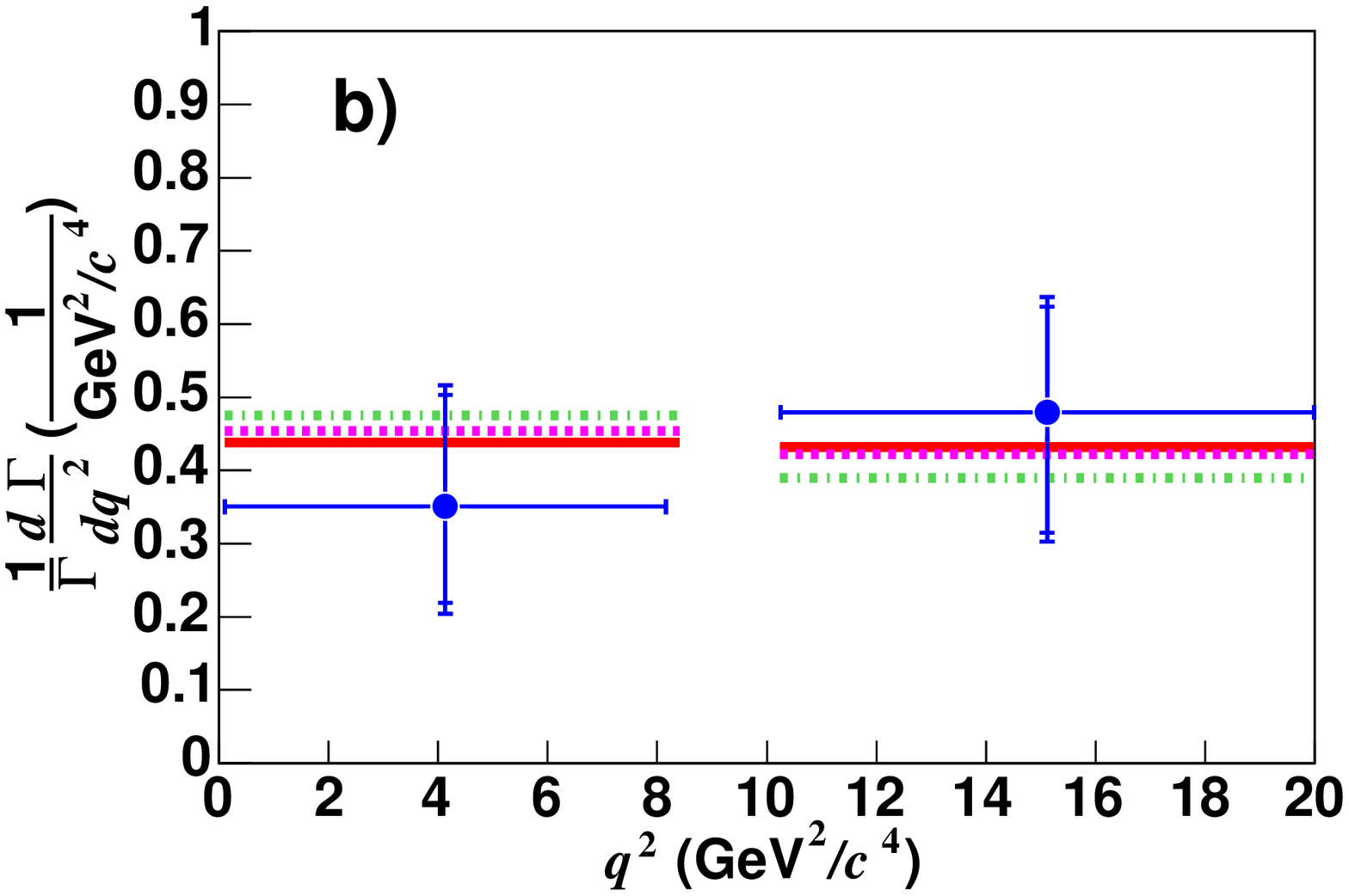}
\caption{Partial branching fractions in bins of $q^2$ for (a) \modekavgll 
and (b) \modekstll, normalized to the total measured branching fraction. 
The points with error bars are data, the lines represent the central values of 
Standard Model predictions based on the form factor models of
Refs.~\cite{bib:TheoryD,bib:TheoryE} (solid lines), 
~\cite{bib:TheoryBc} (dashed lines), and 
~\cite{bib:TheoryBa,bib:ErratumTheoryBa} (dot-dashed lines).}
\label{fig:pbffigs}
\end{center}
\end{figure*}

The partial branching fractions obtained from the fits to 
\mes, \deltaE, and \mkpi in two bins of $q^2$ are shown in 
Table~\ref{tab:bintable}. The results are generally consistent with 
the $q^2$ dependence predicted in recent Standard Model based 
form factor calculations (Figure~\ref{fig:pbffigs}).

\begin{table*}[ntb]
\begin{center} 
\caption[Partial branching fraction results in the combined
$\Kmaybestar\ellell$ decay modes in bins of $q^2$.]  
{ 
Results from fits to the combined $\Kmaybestar\ellell$ decay modes in bins
of $q^2$. The columns from left to right are: fitted $q^2$ range,
partial branching fraction, longitudinal $K^*$ polarization $F_{L}$,
and the lepton forward-backward asymmetry $A_{FB}$.  The first and
second uncertainties are statistical and systematic, respectively. 
In \modekavgll, $A_{FB}$ is measured in the charged $B$ decay modes only.
The constraints for each combined fit 
are described in the text. The partial branching fractions are defined such 
that they include the estimated rate within the vetoed $J/\psi$ and $\psi(2S)$ 
resonance regions where appropriate.}
\label{tab:bintable}
        \begin{tabular*}{\linewidth}{
@{\extracolsep{\fill}}l
@{\extracolsep{\fill}}c
@{\extracolsep{\fill}}c
@{\extracolsep{\fill}}r
}\hline\hline
 & \multicolumn{3}{c}{\modekstll}\\  
$q^2 (\gev^2/c^4)$ & \BR ($10^{-6}$) & $F_L$ & \multicolumn{1}{c}{$A_{FB}$}\\ \hline \vspace{-.1in}\\\vspace{.04in}
\vspace{.04in} $0.1 - 8.41$ 	& $0.27^{+0.12}_{-0.10} \pm 0.05$ & $0.77^{+0.63}_{-0.30} \pm 0.07$& $> 0.19$ (95\%CL)\\ 
\vspace{.04in} $> 10.24$ 		& $0.37^{+0.13}_{-0.11} \pm 0.05$ & $0.51^{+0.22}_{-0.25} \pm 0.08$ & $0.72^{+0.28}_{-0.26} \pm 0.08$ \\ \hline \vspace{-.1in}\\\vspace{.04in}  
\vspace{.04in} $> 0.1$ 		& $0.73^{+0.20}_{-0.18}\pm0.11$ & $0.63^{+0.18}_{-0.19} \pm 0.05$ & $> 0.55$ (95\%CL)\\  \\
 & \multicolumn{3}{c}{\modekavgll}\\ 
$q^2 (\gev^2/c^4)$ & \BR ($10^{-6}$)& $F_S$ & \multicolumn{1}{c}{$A_{FB}$}\\ \hline \vspace{-.1in}\\\vspace{.04in}  
\vspace{.04in} $0.1 - 8.41$ 	& $0.10^{+0.04}_{-0.04} \pm 0.01$ & 0 & $-0.49^{+0.51}_{-0.99} \pm 0.18$ \\ 
\vspace{.04in} $> 10.24$ 		& $0.22^{+0.05}_{-0.05} \pm 0.02$ & 0 & $0.26^{+0.23}_{-0.24} \pm 0.03$ \\ \hline \vspace{-.1in}\\\vspace{.04in}
\vspace{.04in} $> 0.1$ 		& $0.34^{+0.07}_{-0.07}\pm0.02$ & $0.81^{+0.58}_{-0.61} \pm 0.46$ & $0.15^{+0.21}_{-0.23} \pm 0.08$ \\ 
 \hline\hline
\end{tabular*}
\end{center} 
\end{table*}

\subsection{$K^*$ polarization}

The fit projections for the \ctk distribution in bins of $q^2$ are shown in 
Figure~\ref{fig:ctkfits} of Appendix~\ref{sec:appendix}. The resulting values 
for the fraction of longitudinal polarization $F_{L}$ are listed in 
Table~\ref{tab:bintable}. Combining all events with $q^2 > 0.1 \gev^2/c^4$, we 
find 
$${F_{L}}{(\modekstll)}_{(q^2 > 0.1 \gev^2/c^4)} = 0.63^{+0.18}_{-0.19} \pm 0.05,$$
where the first error is statistical, and the second systematic.

The measured values of $F_{L}$ are consistent with the SM expectation in both 
$q^2$ ranges (Figure~\ref{fig:afbfig}) and integrated over all 
$q^2 > 0.1 \gev^2/c^4$. However, the large statistical uncertainties do not
allow the determination of the sign of $C_{7}$ from this measurement at 
present.

\subsection{Lepton forward-backward asymmetry}

The fit projections for the \ctl distribution in the 
\modekll mode are shown in Figure~\ref{fig:ctlkllfits} of 
Appendix~\ref{sec:appendix}.  Combining all events with $q^2 > 0.1 \gev^2/c^4$,
we find for the \modekll mode
$${A_{FB}}{(\modekll)}_{(q^2 > 0.1 \gev^2/c^4)} = 0.15^{+0.21}_{-0.23} \pm 0.08,$$
$${F_S}{(\modekll)}_{(q^2 > 0.1 \gev^2/c^4)} = 0.81^{+0.58}_{-0.61} \pm 0.46,$$
where the first errors are statistical, and the second systematic.  The 
correlation coefficient between these two meaurements is +0.23.  
Both $A_{FB}$ and $F_{S}$ are consistent with the SM prediction of zero.
As a cross-check, we have also performed similar fits in the low and high
$q^2$ regions for $A_{FB}$, where due to limited statistics $F_S$ must be 
fixed to zero; the resulting asymmetries
are $-0.49^{+0.51}_{-0.99}\pm0.18$ and $0.26^{+0.23}_{-0.24}\pm0.03$, 
respectively, which again are both consistent with zero asymmetry.

\begin{figure*}[ntb!]
\begin{center}
\includegraphics[width=0.49\linewidth]{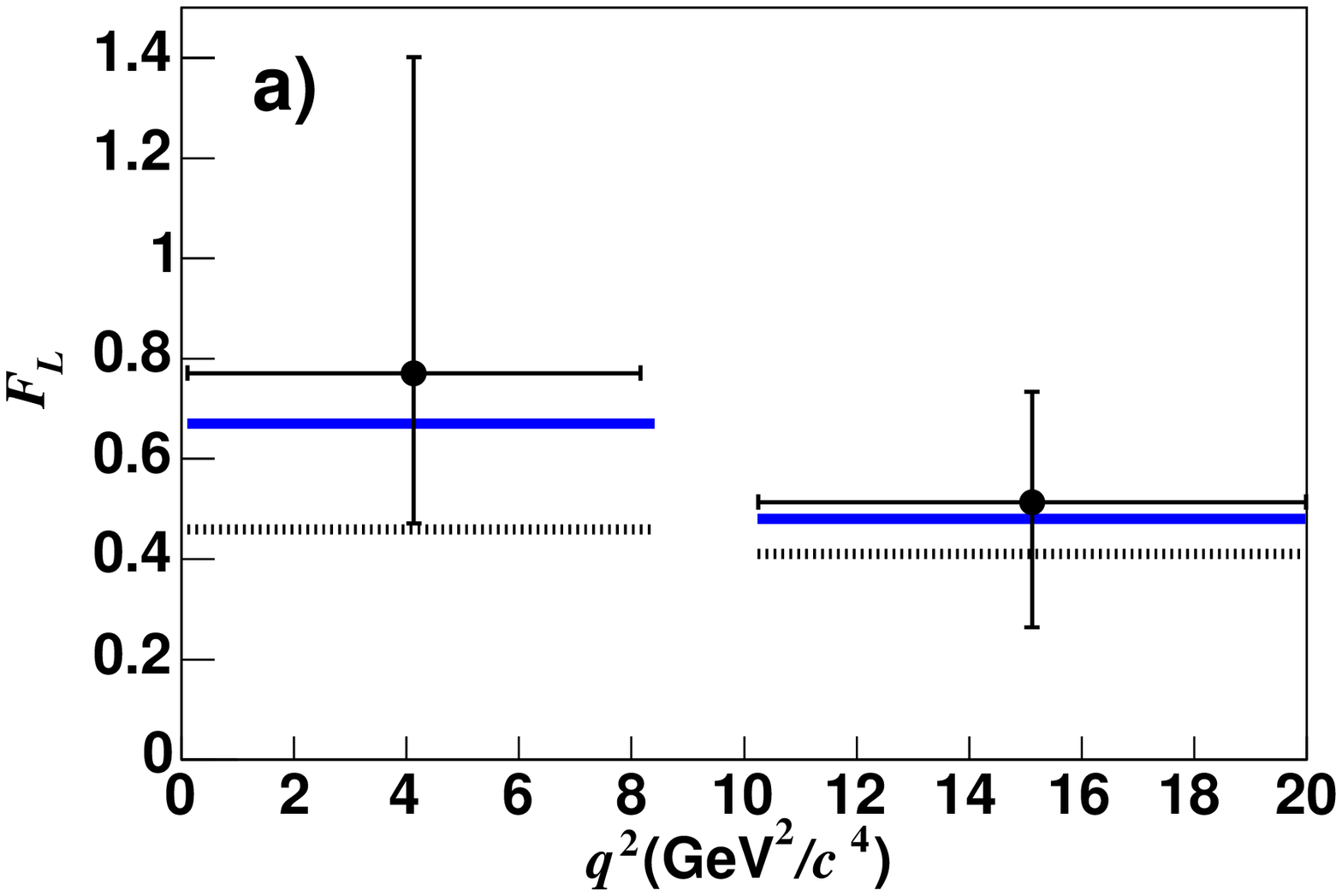}
\includegraphics[width=0.49\linewidth]{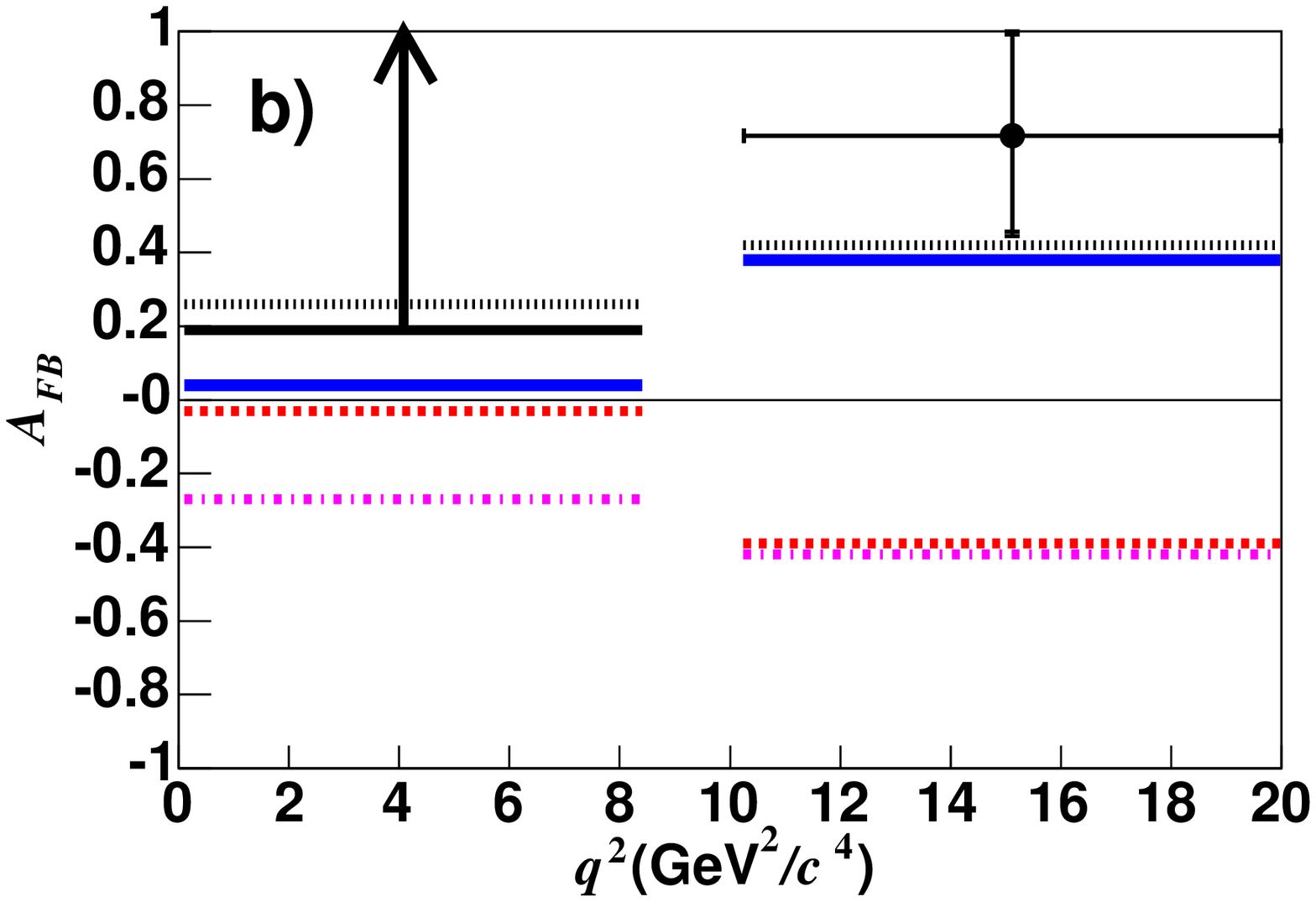}
\caption{(a) $F_{L}(q^{2})$ and (b) $A_{FB}(q^{2})$ in \modekstll. 
The points with error bars are data, with the 
arrow at low $q^2$ in $A_{FB}$ indicating the $95\%$ CL allowed region. 
The lines represent the predictions of 
the SM (solid lines), $C_{7}^{\rm eff} = -C_{7}(SM)$ (dotted lines), 
$C_{9}^{\rm eff}C_{10}^{\rm eff} = -C_{9}C_{10}(SM)$ (dashed lines), and 
$C_{7}^{\rm eff},C_{9}^{\rm eff}C_{10}^{\rm eff} = -C_{7}(SM),-C_{9}C_{10}(SM)$
(dot-dashed lines) with the form factor model of Ref.~\cite{bib:TheoryE}. 
In the case of $F_{L}$, the two solutions with 
$C_{9}^{\rm eff}C_{10}^{\rm eff} = -C_{9}C_{10}(SM)$ are not displayed; 
they are nearly identical to the two shown. }
\label{fig:afbfig}
\end{center}
\end{figure*}

The fit projections for the \ctl distribution in the \modekstll mode are 
shown in Figure~\ref{fig:ctlfits} of Appendix~\ref{sec:appendix}, and the 
resulting values of $A_{FB}$ listed in Table~\ref{tab:bintable}. We find a 
large positive asymmetry in the high $q^2$ region, consistent with the SM 
expectation. This disfavors new physics scenarios in which the product of 
the $C^{\rm eff}_{9}$ and $C^{\rm eff}_{10}$ Wilson coefficients have the 
same magnitude but opposite relative sign as in the SM, which would result 
in a large negative asymmetry at high $q^2$ (Figure~\ref{fig:afbfig}). 

For the low $q^2$ region and the region integrated over all $q^2 > 0.1
\gev^2/c^4$, the $A_{FB}$ value corresponding to the maximum
likelihood is positive, but is near the boundary at which a larger
$A_{FB}$ will result in a negative, undefined value for the extended
likelihood function.  For these maximally asymmetric cases the
$A_{FB}$ result is computed as a one-sided lower limit using a toy
Monte Carlo method.  For fixed values of $A_{FB}$, we randomly
generate from the experimentally measured PDFs an ensemble of toy
experiments, and find the value of $A_{FB}$ for which $5\%$ of
experiments in the ensemble have a maximium likelihood fit resulting
in a maximally positive $A_{FB}$.  The uncertainties in the other PDF
parameters are accounted for by varying them randomly for each
generated experiment in the ensemble according to normal distributions
determined by the parameters' measured central values and uncertainties.  We
account for systematic uncertainties that do not correspond to continuous
PDF parameters, such as the
choice of combinatorial background PDFs for \ctl, by
generating ensembles for each PDF variation and choosing that which
results in the lowest lower limit.  With this method, we find $A_{FB}
> 0.19$ at 95\% CL for the low $q^2$ region.  Combining all events
with $q^2 > 0.1 \gev^2/c^4$, we find for the \modekstll mode at $95\%$
CL
$${A_{FB}}{(\modekstll)}_{(q^2 > 0.1 \gev^2/c^4)} > 0.55.$$
The corresponding fit projections shown in Figure~\ref{fig:ctlfits} are 
produced by fixing the $A_{FB}$ of the signal component to its maximum 
physical value.

\subsection{Search for lepton flavor-violation}

\begin{figure}[ntb!]
\begin{center}
\includegraphics[width=1.0\linewidth]{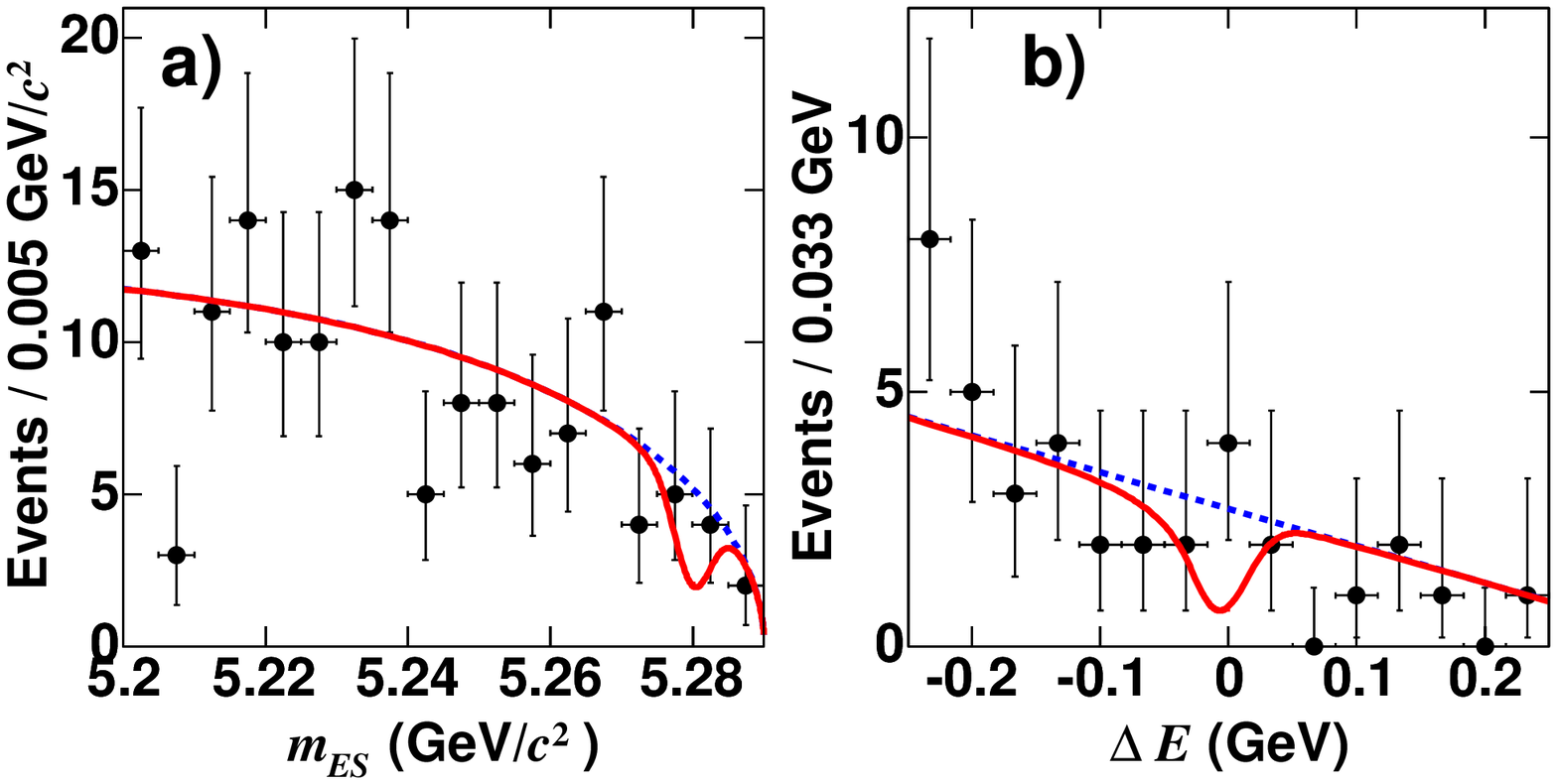}
\caption{
Distributions of the fit variables in $Ke\mu$ data (points),
compared with projections of the combined fit (curves): (a) $m_{\rm
ES}$ distribution after requiring $-0.11<\Delta E<0.05\ {\rm GeV}$ and
(b) $\Delta E$ distribution after requiring
$|m_{\rm ES} - m_{B}| < 6.6\ {\rm MeV}/c^2$.
The solid curve is the sum of all fit components,
including signal; the dashed curve is the sum of all background
components.
}
\label{fig:kemfig}
\end{center}
\end{figure}

\begin{figure}[ntb!]
\begin{center}
\includegraphics[width=1.0\linewidth]{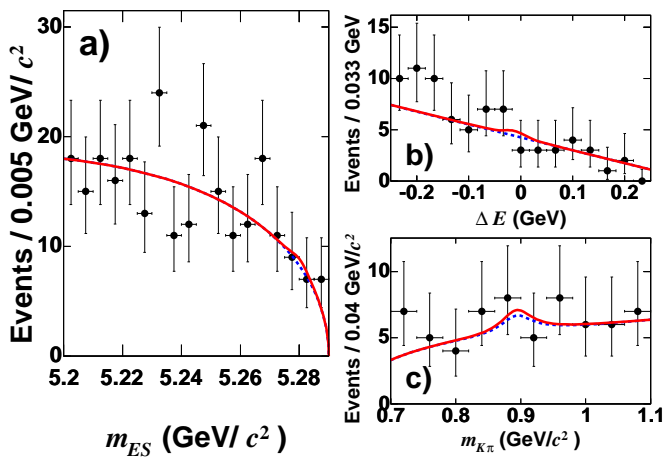}
\caption{
Distributions of the fit variables in $K^*e\mu$ data (points),
compared with projections of the combined fit (curves):
(a) \mes after requiring $-0.11<\Delta E<0.05\ {\rm GeV}$
and $0.817< \mkpi <0.967\ {\rm GeV}/c^2$,
(b) $\Delta E$ after requiring
$|m_{\rm ES} - m_{B}| < 6.6\ {\rm MeV}/c^2$,
$0.817< \mkpi <0.967\ {\rm GeV}/c^2$, and
(c) \mkpi after requiring
$|m_{\rm ES} - m_{B}| < 6.6\ {\rm MeV}/c^2$
and $-0.11<\Delta E<0.05\ {\rm GeV}$.
The solid curve is the sum of all fit components, including signal; the
dashed curve is the sum of all background components.
}
\label{fig:kstemfig}
\end{center}
\end{figure}

We extract the signal yield in the \modekavgem and \modekstem final states
in a similar manner as the $\Kmaybestar\ellell$ decays, with the particle
identification requirements modified to select $e^{\pm}\mu^{\mp}$ pairs.
The signal efficiencies for these modes are determined from simulations
where the $B$ decays according to a simple three-particle phase space model. 
The results are shown in Table~\ref{tab:lfvlimits}. As any physics that 
allows these decays will not necessarily affect the $e^{+}\mu^{-}$ and 
$e^{-}\mu^{+}$ states equally, we quote results for each charge state
in addition to combined charge-averaged results. The projections of the data 
overlayed with the results of the combined fits are shown in 
Figures~\ref{fig:kemfig} and ~\ref{fig:kstemfig}. We find no evidence for a 
signal in any of these channels, and therefore set upper limits on these 
processes. For the combined lepton-charge averaged, $B$-charge averaged modes 
we find
$${\cal B}(B \rightarrow Ke\mu) < 3.8 \times 10^{-8},$$
$${\cal B}(B \rightarrow K^{*}e\mu) < 51 \times 10^{-8},$$
at $90\%$ CL. These limits are significantly more stringent than those 
of previous searches~\cite{bib:cleolfv, bib:babarlfv}.

\begin{table}[h]
\begin{center} 
\caption[Results from fits to the combined $\Kmaybestar\ellell$ decay modes.]
 {
Results from fits to lepton flavor-violating decay modes.
The columns from left are: decay mode, fitted signal yield, selection 
efficiency, relative uncertainty on the branching fraction 
due to the systematic error on the efficiency 
estimate, systematic error on the branching fraction introduced by the 
systematic error on the fitted signal yield, and the $90\%$ C.L. limit on the 
branching fraction.  
The constraints for combined fits 
are described in the text.} 
\label{tab:lfvlimits}
 \begin{tabular}{lrrrrrr} \hline \hline
 \multicolumn{1}{c}{Mode}
 & \multicolumn{1}{c}{Yield}
 & \multicolumn{1}{r}{$\epsilon$ (\%)}
 & \multicolumn{1}{c}{\BR  ($10^{-8}$)}
 & \multicolumn{1}{c}{\BR \ UL ($10^{-8}$)}
 \\
 \hline \vspace{-.1in}\\\vspace{.04in}
$K^+ e^+ \mu^-$     & $-3.5^{+2.1}_{-1.4}$ & 12.6 & $-12.1^{+7.4}_{-5.0} \pm 2.3$ & $9.1$   \\  \vspace{.04in} 
$K^+ e^- \mu^+$     & $-0.8^{+2.1}_{-1.3}$ & 12.6 & $-2.9^{+7.4}_{-4.4} \pm 1.9$  & $13$  \\  \vspace{.04in}  
$K^+ e\mu$          & $-3.2^{+2.7}_{-1.7}$ & 12.6 & $-11.1^{+9.3}_{-5.9} \pm 3.2$ &  $9.1$   \\  \vspace{.04in} 
$K^0 e\mu$          & $-2.9^{+1.9}_{-1.3}$ & 12.5 & $-30^{+\ \,19}_{-\ \,13}\pm\,\,15$                &  $27$  \\  \vspace{.04in} 
$K^{*0} e^+ \mu^-$  & $1.1^{+3.6}_{-2.1}$  & 10.4 & $7^{+\ \,23}_{-\ \,13} \pm\ \,\,\,5$             &  $53$  \\  \vspace{.04in}  
$K^{*0} e^- \mu^+$  & $-1.1^{+3.5}_{-2.2}$ & 10.4 & $-7^{+\ \,22}_{-\ \,14}\pm\ \,\,7$            &  $34$  \\  \vspace{.04in}  
$K^{*0} e\mu$       & $0.9^{+4.6}_{-2.9}$  & 10.4 & $6^{+\ \,\,29}_{-\ \,\,18} \pm\ \,\,9$             &  $58$  \\  \vspace{.04in}  
$K^{*+} e^+ \mu^-$  & $0.4^{+3.4}_{-2.3}$  & 10.0 & $9^{+\ \,65}_{-\ \,44} \pm \,22$                  &  $130$ \\  \vspace{.04in}  
$K^{*+} e^- \mu^+$  & $-1.7^{+3.3}_{-2.0}$ & 10.0 & $-32^{+\ \,63}_{-\ \,38} \pm \,15$                &  $99$  \\  \vspace{.04in}  
$K^{*+} e\mu$       & $-0.2^{+4.2}_{-3.1}$ & 10.0 & $-4^{+\ \,80}_{-\ \,59} \pm \,32$                 &  $140$ \\ \hline  \vspace{-.1in}\\\vspace{.04in}
$K e\mu$             & $-4.9^{+2.9}_{-1.9}$ & -   & $-12.1^{+7.0}_{-4.6} \pm 3.0$ &  $3.8$   \\  \vspace{.04in}  
$K^{*} e\mu$         & $1.0^{+5.5}_{-3.7}$  & -   & $48^{+\ \,26}_{-\ \,17} \pm\ 11$              &  $51$      \\ \hline \hline  
\end{tabular}
\end{center} 
\end{table}

\section{Conclusions} \label{sec:conclusions}

We have measured the branching fractions, partial branching fractions, 
direct $\CP$ asymmetries, ratio of muons to electrons, fraction of 
longitudinal $K^*$ polarization, and lepton forward-backward asymmetries 
in the rare FCNC decays \modekavgll and \modekstll.

The branching fraction, $A_{\CP}$, $R_{K}$, and $F_{L}$ results are
all consistent with the Standard Model predictions for these decays.
The values of $A_{FB}$ and the scalar contribution $F_{S}$ measured 
in the \modekll channel are consistent with the expected value of zero.  
In the \modekstll channel the large positive value of $A_{FB}$ at high $q^2$ is
consistent with the SM and disfavors new physics scenarios in which
the relative sign of the product of the $C_{9}$ and $C_{10}$ Wilson 
coefficients is opposite that of the SM. At low $q^2$ a positive value 
of $A_{FB}$ is also favored, with a $95\%$ CL lower limit that is slightly 
above the SM prediction, as derived using the form factor models of
Refs.~\cite{bib:TheoryBc,bib:TheoryE}.

In addition, we have obtained upper limits on the lepton flavor-violating 
decays $B \rightarrow Ke\mu$ and $B \rightarrow K^{*}e\mu$ that are 
approximately one order of magnitude lower than those of previous 
searches.

We note that the Belle collaboration has recently reported~\cite{bib:belleafb}
a measurement of the integrated forward-backward asymmetries, finding 
$\bar{A}_{FB}(\modekll) = 0.10 \pm 0.14 \pm 0.01$ and 
$\bar{A}_{FB}(\modekstll) = 0.50 \pm 0.15 \pm 0.02$. From a fit 
to the \ctl and $q^2$ distributions, they conclude that scenarios in which 
the product of $C_{9}$ and $C_{10}$ has the opposite sign as expected in the 
SM are disfavored, consistent with the results reported here.

All of the measurements reported here are limited by statistical 
uncertainties, and can be improved with the addition of more data.

\section{Acknowledgments} \label{sec:ack}


\input pubboard/acknow_PRL


\clearpage
\newpage
\appendix
\section{Fits to angular distributions} \label{sec:appendix}

\begin{figure*}[htbp]
\begin{center}
\includegraphics[width=0.32\linewidth]{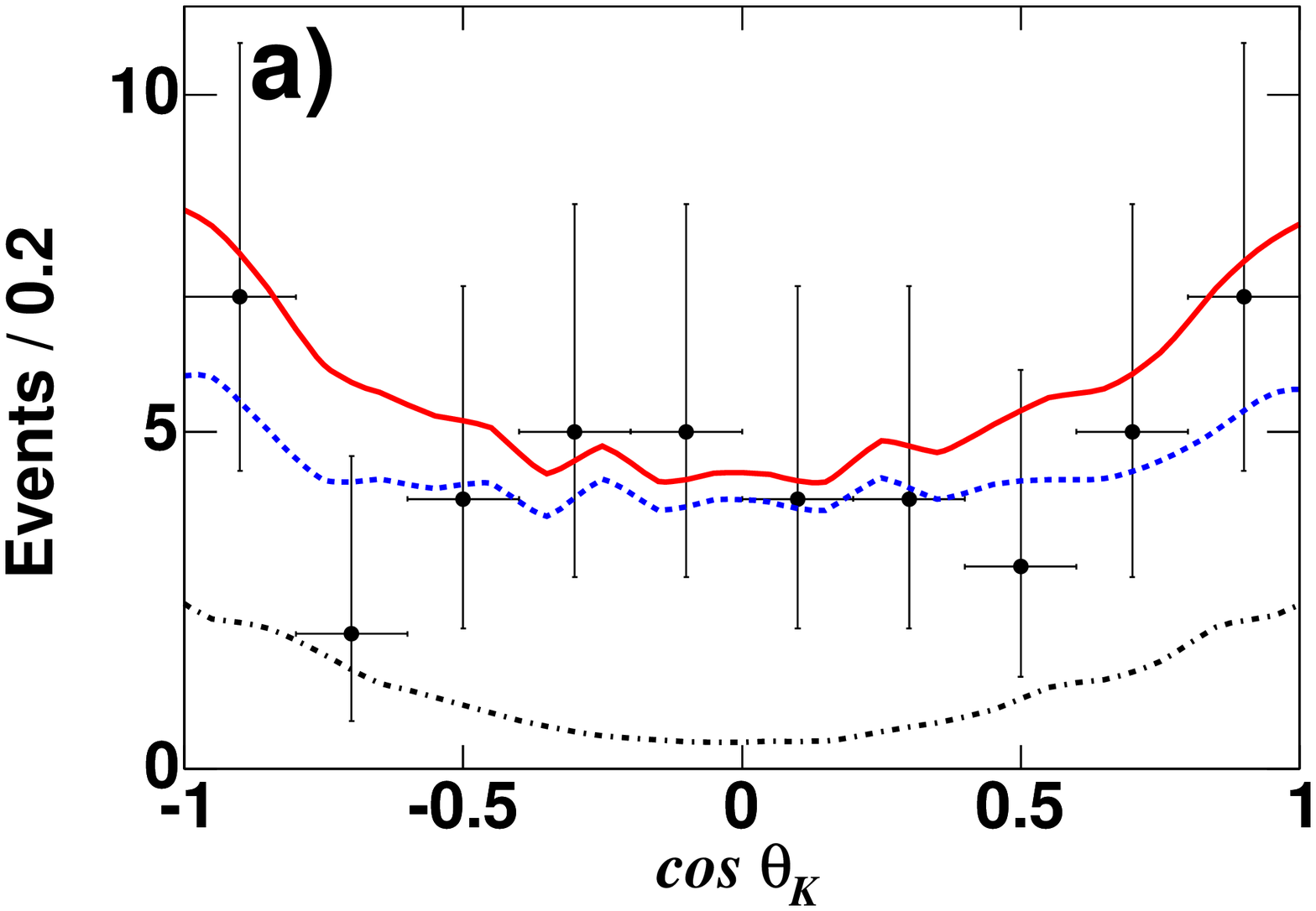}
\includegraphics[width=0.32\linewidth]{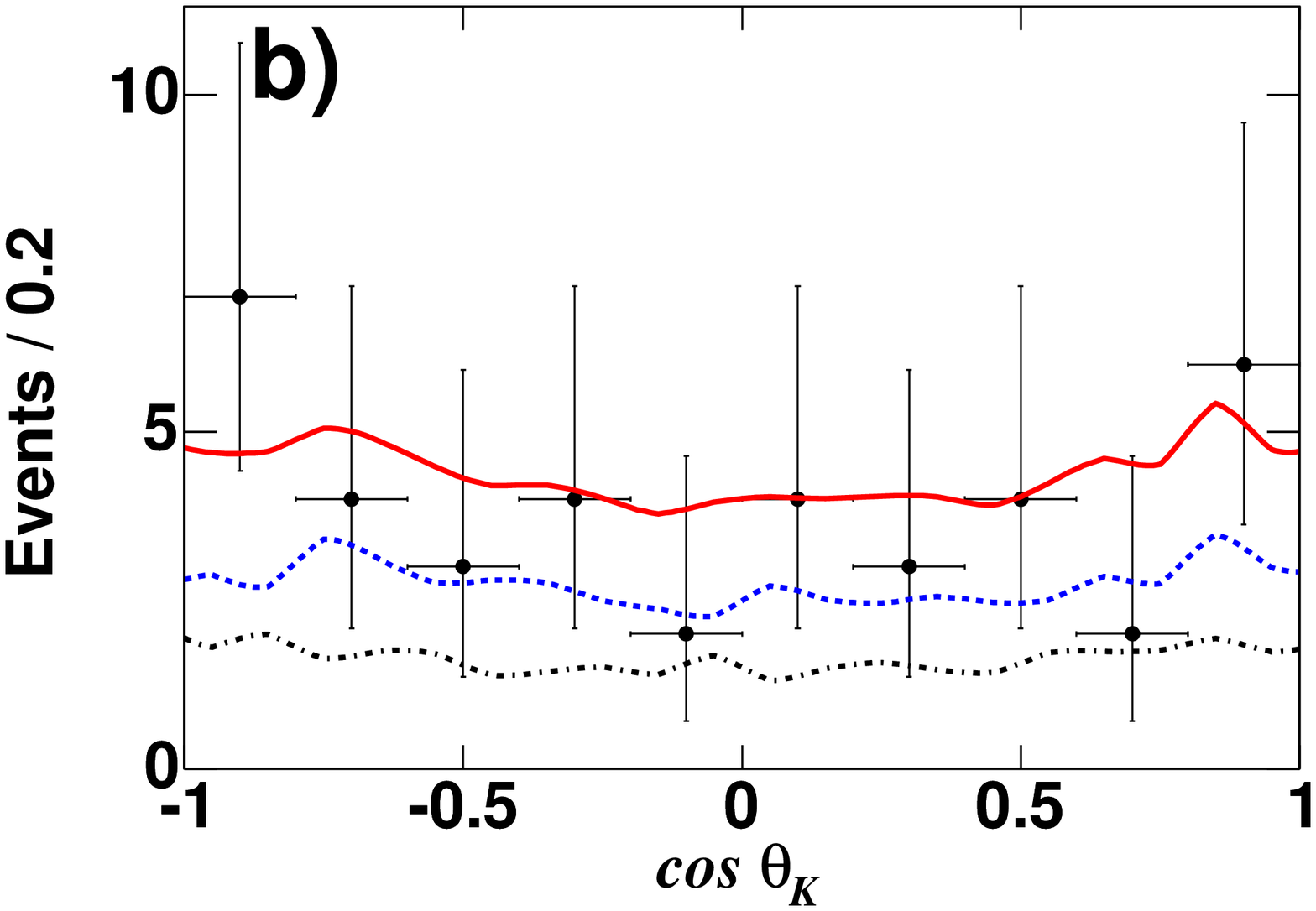}
\includegraphics[width=0.32\linewidth]{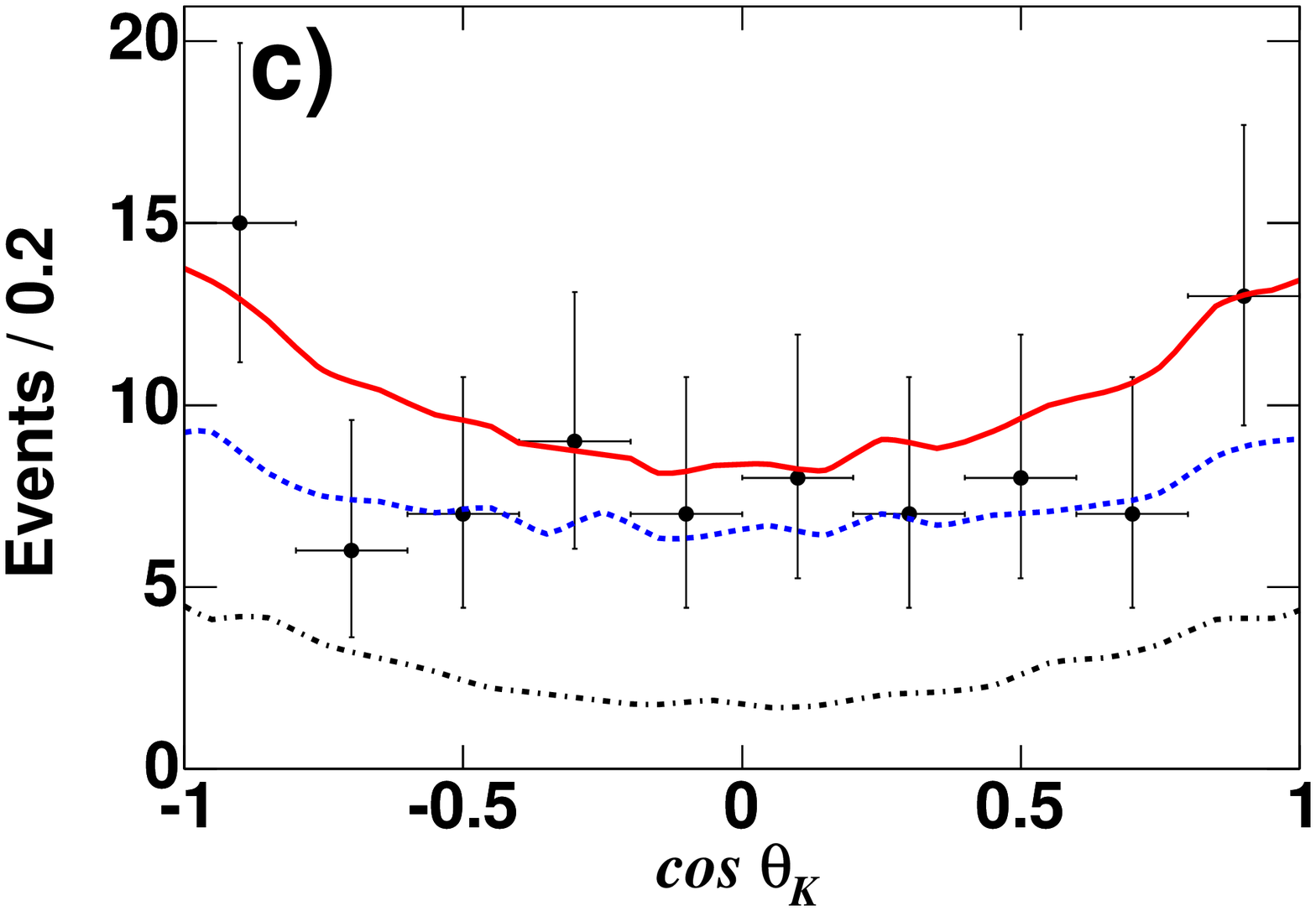}
\caption{Distributions of the fit variable \ctk in \modekstll data (points),
compared with projections of the combined fit (curves) after requiring 
$-0.11<\Delta E<0.05\ {\rm GeV}$, $|m_{\rm ES} - m_{B}| < 6.6\ {\rm MeV}/c^2$, 
and $0.817< \mkpi <0.967\ {\rm GeV}/c^2$.
The solid curve is the sum of all fit components, the dashed curve is the sum 
of all background components, and the dot-dashed curve is the signal component. The $q^2$ regions (a) $0.1 < q^2 < 8.41 \gev^2/c^4$, (b) $q^2 > 10.24 \gev^2/c^4$, and (c) $q^2 > 0.1 \gev^2/c^4$ are shown.}
\label{fig:ctkfits}
\end{center}
\end{figure*}

\begin{figure*}[htbp]
\begin{center}
\includegraphics[width=0.32\linewidth]{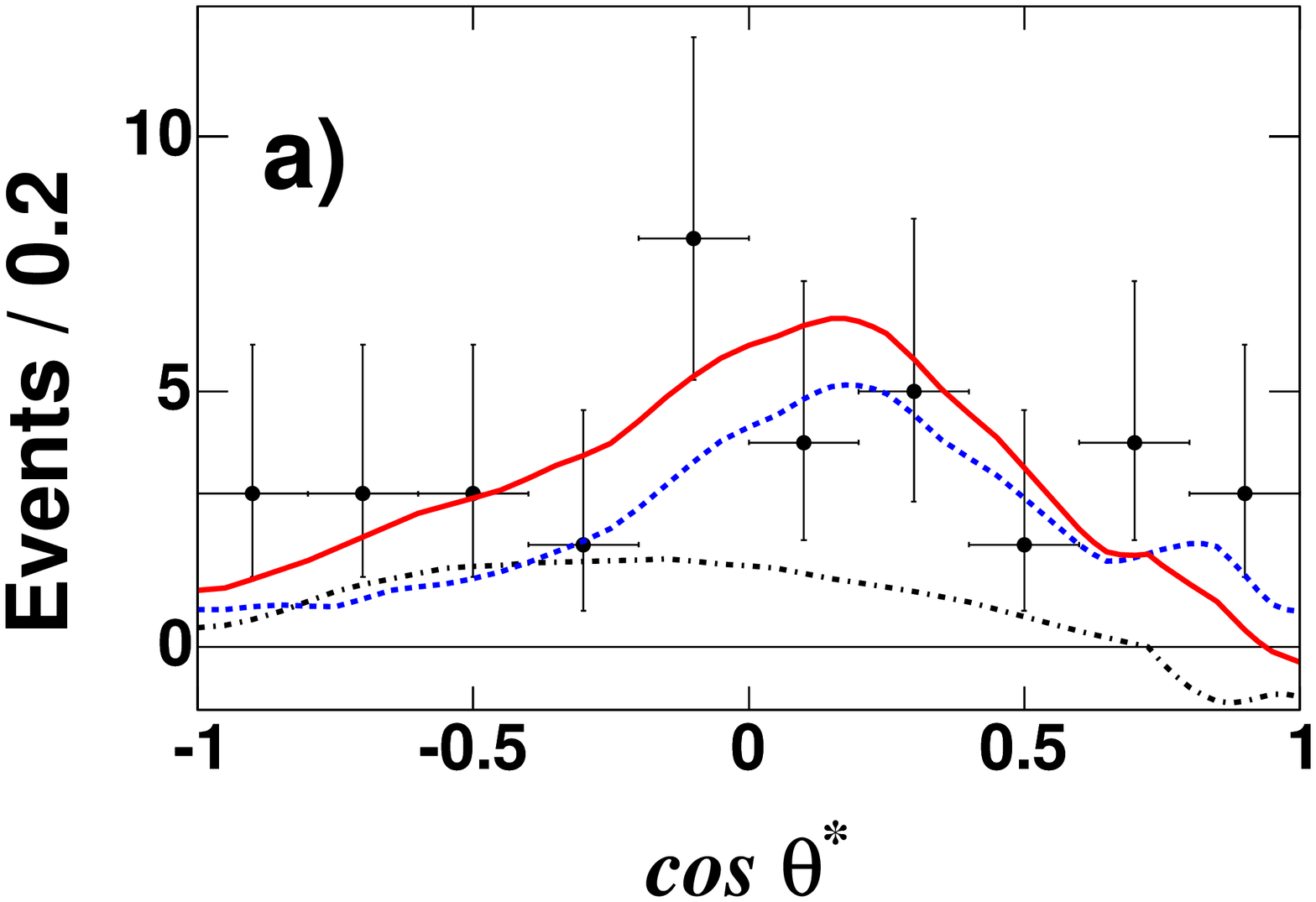}
\includegraphics[width=0.32\linewidth]{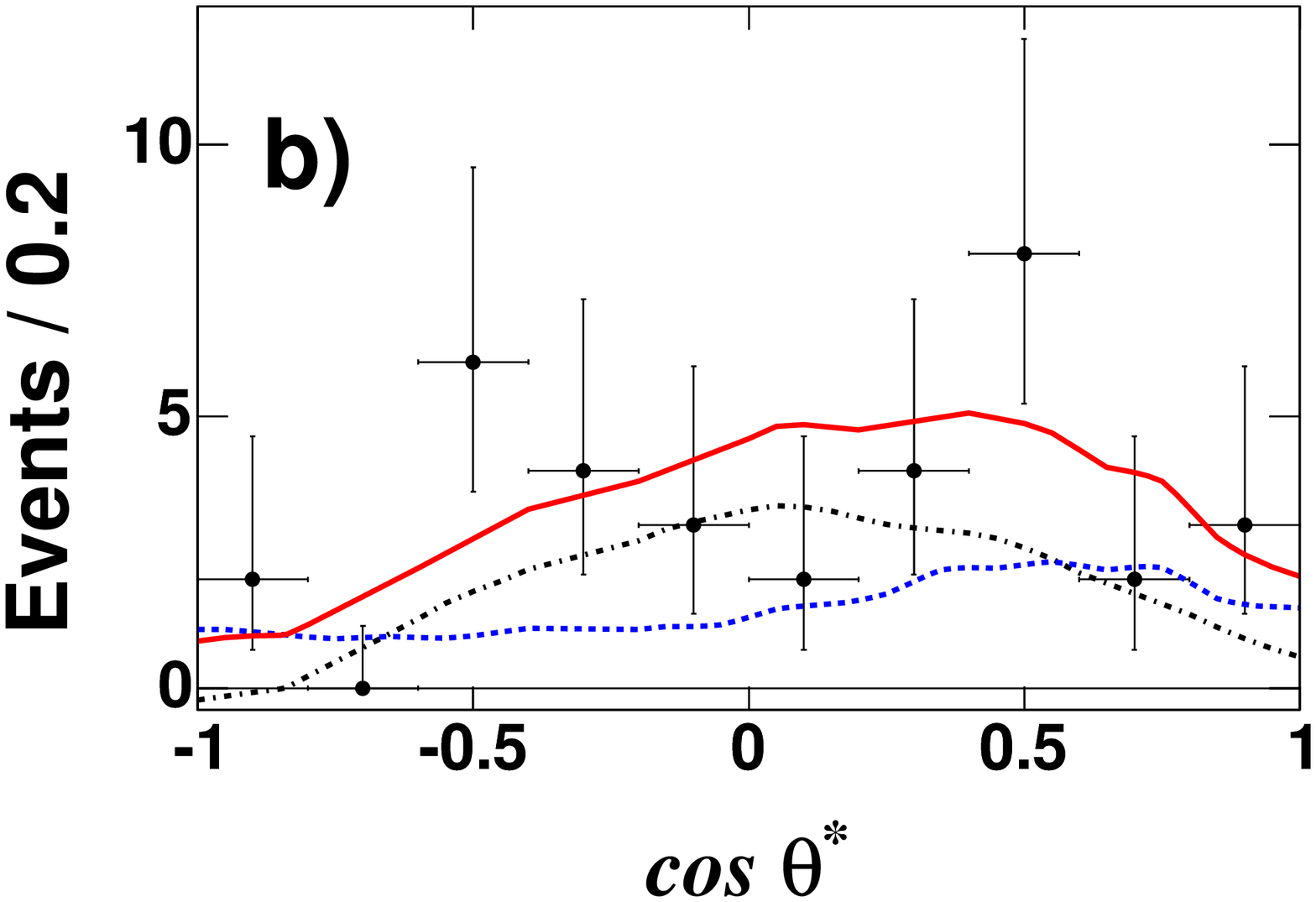}
\includegraphics[width=0.32\linewidth]{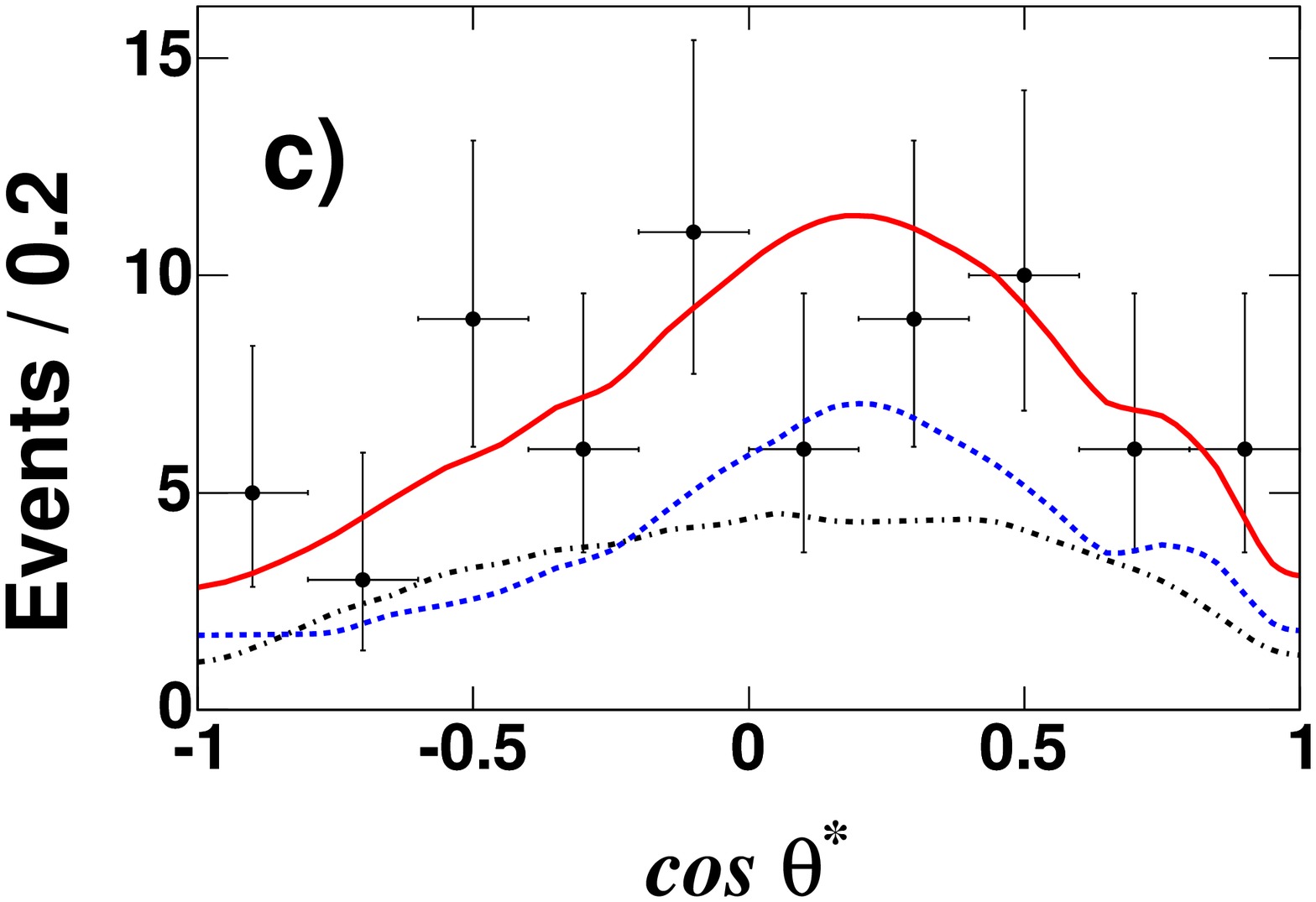}
\caption{Distributions of the fit variable \ctl in \modekll data (points),
compared with projections of the combined fit (curves) after requiring 
$-0.11<\Delta E<0.05\ {\rm GeV}$ and
 $|m_{\rm ES} - m_{B}| < 6.6\ {\rm MeV}/c^2$.
The solid curve is the sum of all fit components, the dashed curve is the sum 
of all background components, and the dot-dashed curve is the signal component.
The $q^2$ regions (a) $0.1 < q^2 < 8.41 \gev^2/c^4$, (b) $q^2 > 10.24 \gev^2/c^4$, and (c) $q^2 > 0.1 \gev^2/c^4$ are shown.  The combined fits shown for (a) and (b) are performed by fixing $F_S$ to zero.}
\label{fig:ctlkllfits}
\end{center}
\end{figure*}

\begin{figure*}[htbp]
\begin{center}
\includegraphics[width=0.32\linewidth]{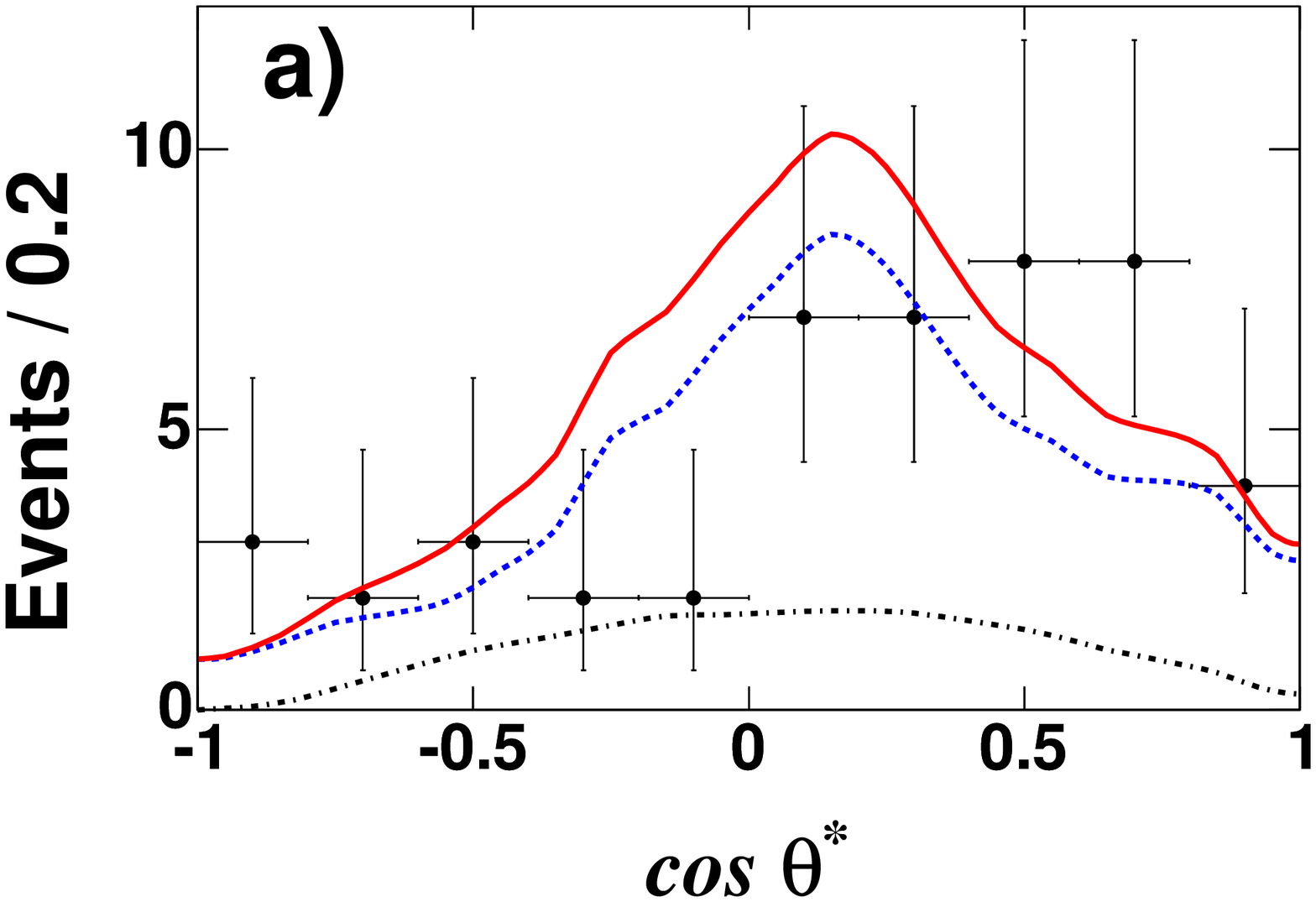}
\includegraphics[width=0.32\linewidth]{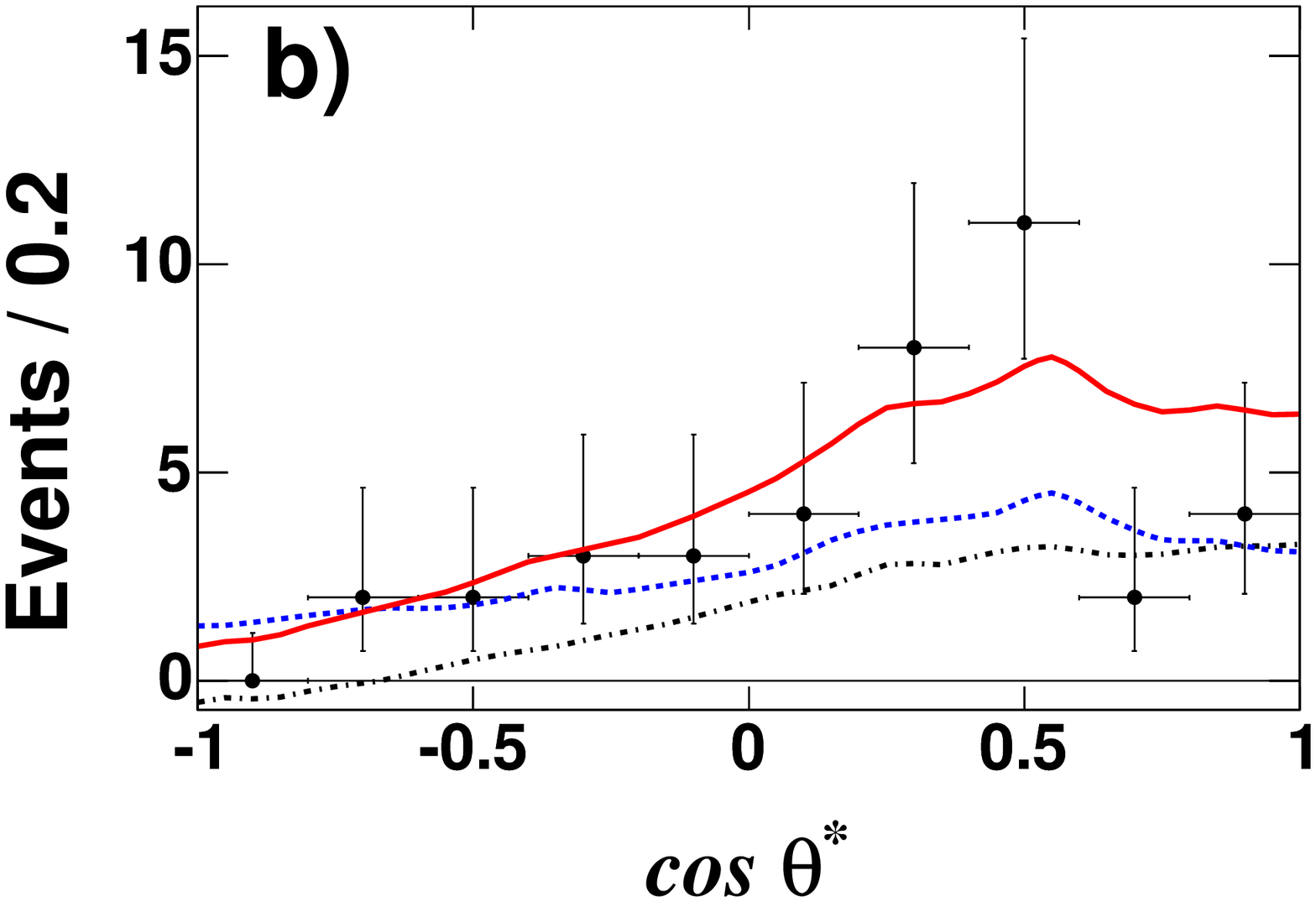}
\includegraphics[width=0.32\linewidth]{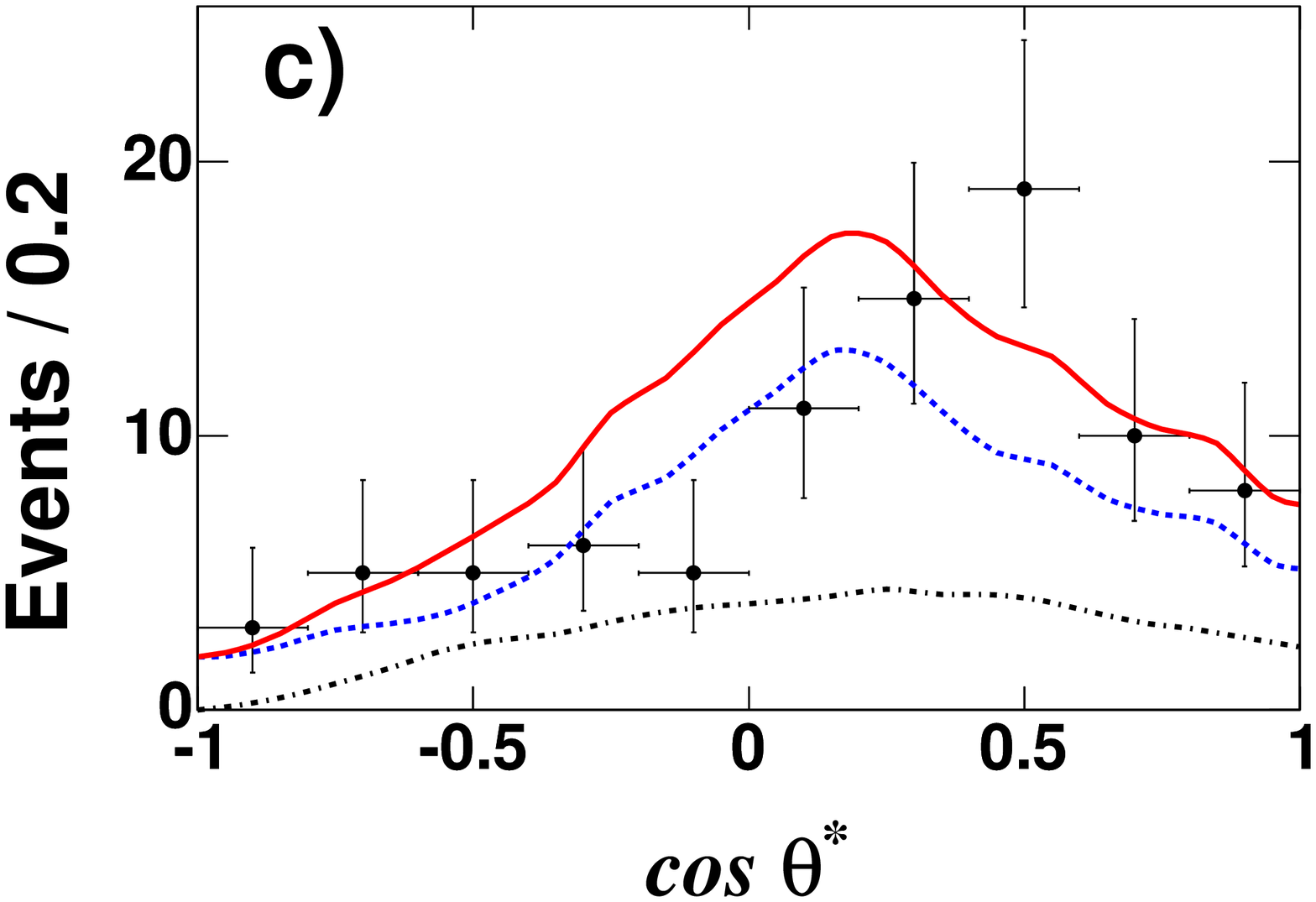}
\caption{Distributions of the fit variable \ctl in \modekstll data (points),
compared with projections of the combined fit (curves) after requiring 
$-0.11<\Delta E<0.05\ {\rm GeV}$, $|m_{\rm ES} - m_{B}| < 6.6\ {\rm MeV}/c^2$, 
and $0.817< \mkpi <0.967\ {\rm GeV}/c^2$.
The solid curve is the sum of all fit components, the dashed curve is the sum 
of all background components, and the dot-dashed curve is the signal component.
The $q^2$ regions (a) $0.1 < q^2 < 8.41 \gev^2/c^4$, (b) $q^2 > 10.24 \gev^2/c^4$, and (c) $q^2 > 0.1 \gev^2/c^4$ are shown.  The combined fits shown for 
(a) and (c) are performed by fixing $A_{FB}$ to its maximal physical value.}
\label{fig:ctlfits}
\end{center}
\end{figure*}

In this appendix we present plots of the \ctk and \ctl distributions in data, 
together with the projections of the combined fits used to extract $F_{L}$ 
and $A_{FB}$. Figure~\ref{fig:ctkfits} shows the fitted \ctk distributions 
for each of the $q^2$ bins considered in this analysis. 
Figures~\ref{fig:ctlkllfits} and~\ref{fig:ctlfits} display the fitted 
\ctl distributions for each of the $q^{2}$ ranges for the \modekll and 
\modekstll decay modes, respectively. For the fits to the \ctl distributions 
in the \modekstll mode, the $K^*$ polarization $F_{L}$ is fixed to its 
measured value, as described in the text. The deviations from a 
smooth parabolic shape in the signal component are the result of the efficiency
and acceptance corrections, which are described by non-parametric histogram 
PDFs. 

\end{document}

%% file: pubboard/authors_feb2006.tex
%
\author{B.~Aubert}
\author{R.~Barate}
\author{M.~Bona}
\author{D.~Boutigny}
\author{F.~Couderc}
\author{Y.~Karyotakis}
\author{J.~P.~Lees}
\author{V.~Poireau}
\author{V.~Tisserand}
\author{A.~Zghiche}
\affiliation{Laboratoire de Physique des Particules, F-74941 Annecy-le-Vieux, France }
% ...
\collaboration{The \babar\ Collaboration}
\noaffiliation

%% file: pubboard/acknow_PRL.tex
We are grateful for the excellent luminosity and machine conditions
provided by our \pep2\ colleagues, 
and for the substantial dedicated effort from
the computing organizations that support \babar.
The collaborating institutions wish to thank 
SLAC for its support and kind hospitality. 
This work is supported by
DOE
and NSF (USA),
NSERC (Canada),
IHEP (China),
CEA and
CNRS-IN2P3
(France),
BMBF and DFG
(Germany),
INFN (Italy),
FOM (The Netherlands),
NFR (Norway),
MIST (Russia), and
PPARC (United Kingdom). 
Individuals have received support from CONACyT (Mexico), 
Marie Curie EIF (European Union),
the A.~P.~Sloan Foundation, 
the Research Corporation,
and the Alexander von Humboldt Foundation.